\documentclass{emulateapj}
\usepackage{lscape}

\def\etal{{\it et~al.}~}
\def\wprime{W$^{\prime}$~}
\def\kms{{km~s$^{-1}$}}
\def\hst{{\it HST}}
\def\vi{\ifmmode(V{-}I)\else$(V{-}I)$\fi}
\def\viz{\ifmmode(V{-}I)_0\else$(V{-}I)_0$\fi}

\def\gz{\ifmmode(g_{475}{-}z_{850})\else$(g_{475}{-}z_{850})$\fi}
\def\gzz{\ifmmode(g_{475}{-}z_{850})_0\else$(g_{475}{-}z_{850})_0$\fi}
\newcommand\lta{\mathrel{\rlap{\lower 3pt\hbox{$\mathchar"218$}}
     \raise 2.0pt\hbox{$\mathchar"13C$}}}
\newcommand\gta{\mathrel{\rlap{\lower 3pt\hbox{$\mathchar"218$}}
     \raise 2.0pt\hbox{$\mathchar"13E$}}}
\def\mbari{\ifmmode\overline{m}_I\else$\overline{m}_I$\fi}
\def\mbarz{\ifmmode\overline{m}_z\else$\overline{m}_z$\fi}
\def\mbar{\ifmmode\overline{m}\else$\overline{m}$\fi}
\def\Mbar{\ifmmode\overline{M}\else$\overline{M}$\fi}
\def\Mbarz{\ifmmode\overline{M_z}\else$\overline{M}_z$\fi}

\slugcomment{Mei et al. 2007, ApJ, 655, 144}

\shorttitle{Three Dimensional Structure of the Virgo Cluster}
\shortauthors{MEI ET AL. 2007, ApJ, 655, 144}

\begin{document}

\title{The ACS Virgo Cluster Survey. XIII. SBF Distance Catalog and
the Three-Dimensional Structure of the Virgo Cluster\altaffilmark{1}}

\author{Simona Mei\altaffilmark{2,3,4}, 
John P. Blakeslee\altaffilmark{5}, 
Patrick C\^ot\'e\altaffilmark{6},
John L. Tonry\altaffilmark{7},
Michael~J.~West\altaffilmark{8,9},
Laura Ferrarese\altaffilmark{6},
Andr\'es Jord\'an\altaffilmark{10},
Eric W. Peng\altaffilmark{6},
Andr\'e Anthony\altaffilmark{6},
David~Merritt\altaffilmark{11}}
\altaffiltext{1}{Based on observations with the NASA/ESA {\it Hubble
Space Telescope} obtained at the Space Telescope Science Institute,
which is operated by the association of Universities for Research in
Astronomy, Inc., under NASA contract NAS 5-26555.}
\altaffiltext{2}{Department of Physics and Astronomy, Johns Hopkins University, Baltimore, MD 21218; smei@pha.jhu.edu}
\altaffiltext{3}{GEPI, Observatoire de Paris, Section de Meudon, 5 Place J.Janssen, 92195 Meudon Cedex, France}
\altaffiltext{4}{University of Paris 7 Denis Diderot, 2 place Jussieu, 75251 Paris Cedex 05, France}
\altaffiltext{5}{Department of Physics and Astronomy, Washington State University, Pullman, WA, 99164--2814}
\altaffiltext{6}{Herzberg Institute of Astrophysics, National Research Council, 5071 West Saanich Road, Victoria, BC, V9E 2E7, Canada}
\altaffiltext{7}{Institute of Astronomy, University of Hawaii, 2680 Woodlawn Drive, Honolulu, HI 96822}
\altaffiltext{8}{Gemini Observatory, Casilla 603, La Serena, Chile}
\altaffiltext{9}{Department of Physics and Astronomy, University of Hawaii, Hilo, HI 96720}
\altaffiltext{10}{European Southern Observatory, Karl--Schwarzschild--Str. 2, 85748 Garching, Germany}
\altaffiltext{11}{Department of Physics, Rochester Institute of Technology, 54 Lomb Memorial Drive, Rochester, NY 14623}

\begin{abstract}
The ACS Virgo Cluster Survey consists of HST/ACS imaging for 100 early--type galaxies
in the Virgo cluster, observed in the F475W ($\approx$ SDSS $g$) and F850LP ($\approx$
SDSS $z$) filters.
We derive distances for 84 of these galaxies using the method of Surface Brightness Fluctuations
(SBF), present the SBF distance catalog, and use this database to examine the 
three-dimensional distribution of early--type galaxies in the Virgo Cluster.
The SBF distance moduli have a mean (random) measurement error of 0.07~mag (0.5~Mpc), 
or roughly three times better than previous SBF measurements for Virgo Cluster galaxies. Five galaxies
lie at a distance of $d \approx$ 23~Mpc and are members of the \wprime Cloud. The
remaining 79 galaxies have a narrow distribution around our adopted distance 
of $\langle d\rangle =$ 16.5$\pm$0.1 (random mean error) $\pm$ 1.1~Mpc 
(systematic). The $rms$ distance scatter of this sample is $\sigma(d) = 0.6\pm0.1$~Mpc, 
with little  or no dependence on morphological type or luminosity class (i.e., 
$0.7\pm0.1$~Mpc and $0.5\pm0.1$~Mpc for the giants and dwarfs, respectively). 
The back-to-front depth of the cluster measured from our sample of early-type 
galaxies is 2.4$\pm$0.4~Mpc (i.e., $\pm~2\sigma$
of the intrinsic distance distribution). The M87 (Cluster A)
and M49 (Cluster B) subclusters are found to lie at distances of 16.7$\pm$0.2 and
16.4$\pm$0.2~Mpc, respectively. There may be a third subcluster
associated with M86. A weak correlation between velocity and line-of-sight 
distance may be
a faint echo of the cluster velocity distribution not having yet completely virialized.
In three-dimensions, Virgo's early-type galaxies appear
to define a slightly triaxial distribution, with axis ratios of (1:0.7:0.5).
The principal axis of the best-fit ellipsoid is inclined
$\sim$ 20$^{\circ}$--40$^{\circ}$ from the line of sight, while
the galaxies belonging to the \wprime Cloud lie on an axis inclined by $\sim 10^{\circ} -
15^{\circ}$.
\end{abstract}


\def\hst{{\it HST}}

\section{Introduction}

Currently favored models of structure formation predict
that rich clusters are assembled through the accretion, along
predominantly radial filaments, of galaxies and subclusters (e.g., van Haarlem \& van de Weygaert 1993;
Bond, Kofman \& Pogosyan 1996; Springel \etal 2005).
The presence of distinct substructures in many clusters --- identified primarily
through statistical analyses of the positions and line-of-sight velocities of
individual galaxies, or from an analysis of the cluster X-ray morphology --- strongly supports this 
picture (e.g., Geller \& Beers 1982; Dressler \& Shectman 1988; Bird 1994;
West, Jones \& Forman 1995; Oegerle \& Hill 2001) and shows that, even
in nearest clusters, the process of virialization is often incomplete
(Fitchett \& Webster 1987; Merritt 1987; Mellier \etal 1988; Mohr,
Fabricant \& Geller 1993). In principle, accurate distances for individual members
of nearby clusters could be used to map out their three-dimensional structure,
identify substructures, help disentangle their internal dynamics, and shed light
on the processes by which clusters grow and virialize.
In practice, though, the extreme difficulty of measuring distances with the accuracy 
needed to resolve the cluster along the line-of-sight ($\sigma_d \lesssim 1$~Mpc)
has thwarted such efforts in even the nearest clusters.

At a distance of $\approx$ 16.5~Mpc, Virgo is the rich cluster nearest to the Milky
Way. As such, it offers the best hope to resolve the three-dimensional structure
of a rich cluster through distance measurements of individual galaxies. It has long
been recognized from the spatial distribution and radial velocities of Virgo
galaxies that the cluster has a complex distribution with several distinct
components. de Vaucouleurs (1961) first proposed the existence of different
substructures --- which he termed {\it clouds} --- extending far from the main
core. Many subsequent studies helped to characterize the internal structure of
the cluster and the relationship with its large-scale surroundings (e.g., de
Vaucouleurs \& de Vaucouleurs 1973;  Helou, Salpeter \&
Krumm 1979; Tully \& Shaya 1984; Huchra 1985; Tanaka \etal 1985).
\begin{figure*}
\epsscale{1.2}
\centerline{\plottwo{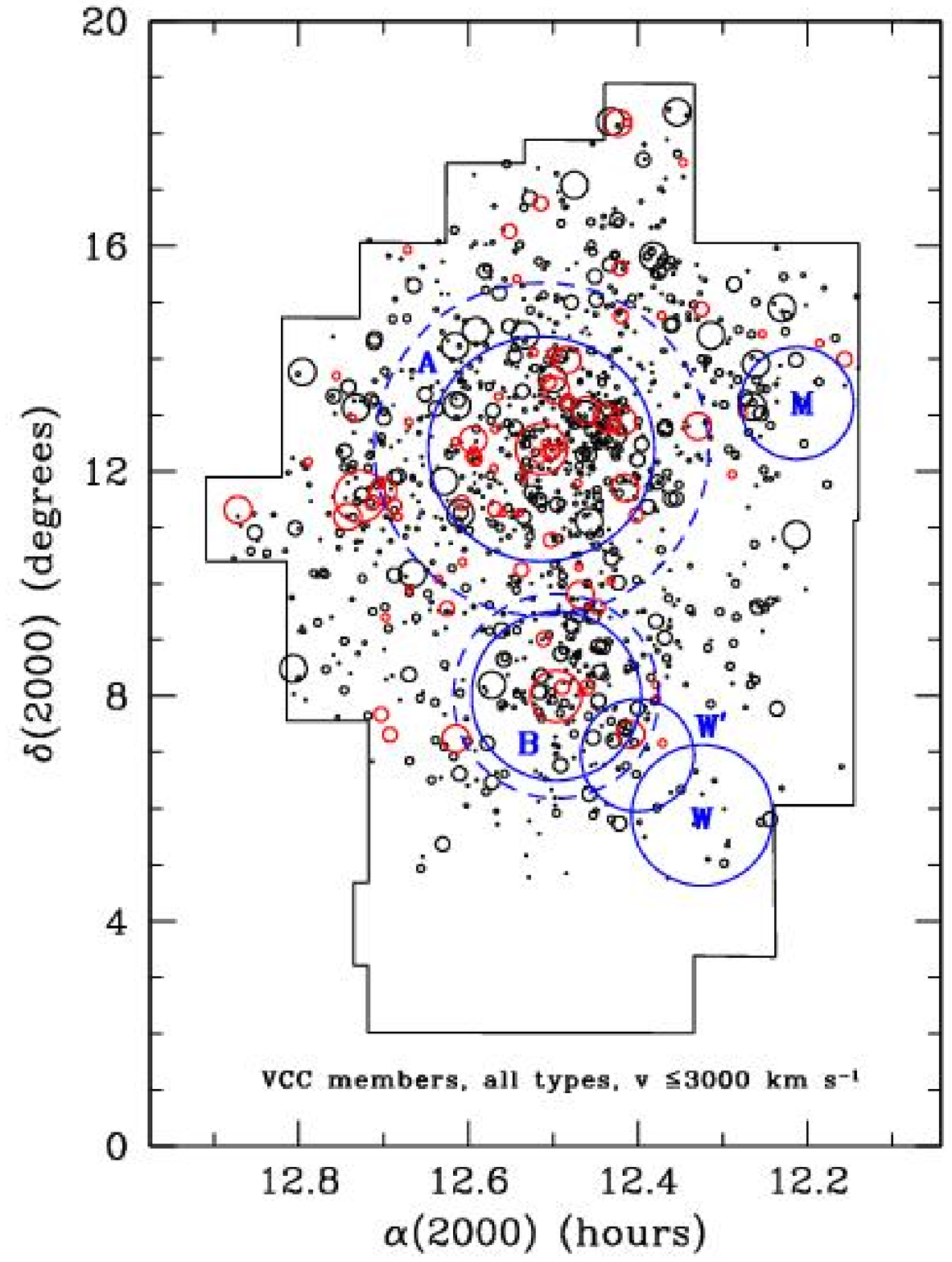}{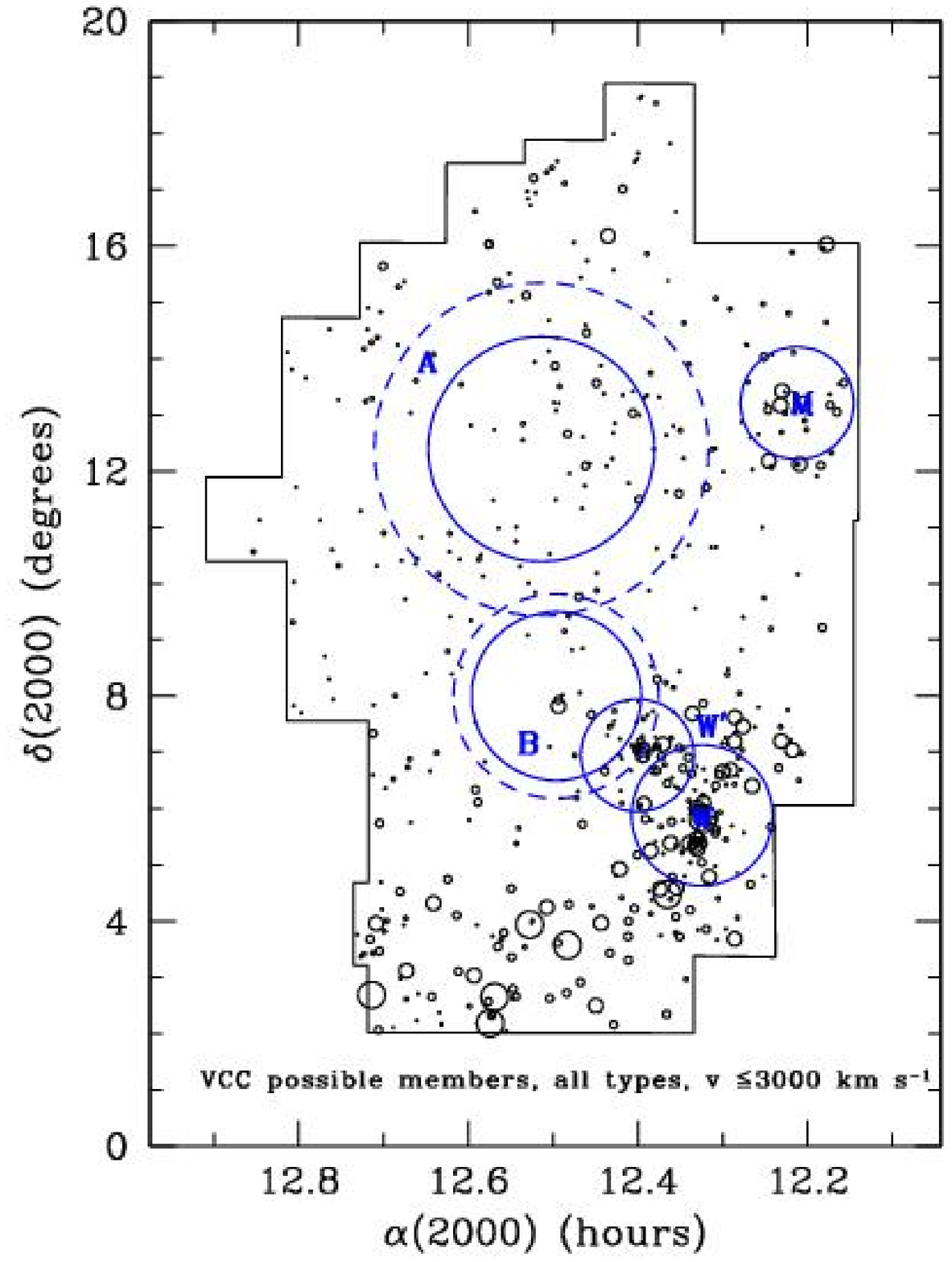}}
\caption{{\it (Left Panel)} The distribution on the sky of galaxies in the Virgo Cluster
Catalog (VCC) classified as {\it members} by Binggeli \etal (1985) and 
excluding galaxies with
heliocentric radial velocities $v_r \ge 3000$ \kms. A total of 1251 galaxies 
are plotted, with no restriction on morphological type. Red
circles indicate the 100 early-type galaxies from the ACS Virgo Cluster Survey.
In all cases, symbol size is proportional to blue luminosity.
The solid blue circles show the location and boundaries of the M87 (A) and 
M49 (B) clusters --- and the M, W$^{\prime}$ and W clouds --- as defined in 
Binggeli \etal (1987). The larger dashed blue
circles centered on M87 and M49 show $r_{200}/2$ for Clusters A and B,
determined from the optical/X-ray mass models from
McLaughlin (1999) and C\^ot\'e \etal (2001, 2003). The solid lines show the boundaries of
the VCC survey, as given in Binggeli \etal (1987).
{\it (Right Panel)} VCC galaxies classified as
{\it possible members} by Binggeli \etal (1985) and having heliocentric radial
velocities of $v_r \le 3000$ \kms for a total of 511 galaxies.
\label{fig1}}
\end{figure*}
To a large extent, the current picture of Virgo's global structure can be traced to the 
wide-field, photographic imaging survey of Binggeli and 
collaborators (e.g., Binggeli, Sandage \& Tammann 1985; Binggeli, Tammann \& Sandage 
1987; Binggeli, Popescu \& Tammann 1993). Galaxy velocities discussed in those papers were collected from previous work (see references in Table~IIa in Binggeli, Sandage \& Tammann 1985 and \S2 in Binggeli, Popescu \& Tammann  1993).
At least six distinct components were adopted in
this series of papers: the main body of the cluster ({\it Cluster A}), which is centered
on the cD galaxy M87 (= NGC4486 = VCC1316), a smaller subcluster ({\it Cluster B}) centered on M49
(= NGC4472 = VCC1226), three compact clouds offset by $\approx 5^{\circ}$ from the main body of
the cluster --- previously named the M, W, and W$^{\prime}$ clouds by de Vaucouleurs (1961) ---
and an elongated Southern Extension (SE) cloud that is visible at declinations of
$\delta \lesssim 5^{\circ}$. The center of the Virgo Cluster as a whole--- as defined
by either the galaxies (Binggeli 1999) or by the X-ray-emitting
gas (B\"ohringer \etal 1994) --- lies close to M87, but is displaced slightly toward
M86 (= NGC4406 = VCC881). Schindler, Binggeli \& B\"ohringer (1999) have concluded
from an optical/X-ray analysis of the cluster that M86 may, like M87 and M49,
define the center of its own distinct subcluster (see also Binggeli 1999).

Most of these studies relied on positions and radial velocities of individual galaxies
to explore the global structure of the cluster. Clearly, accurate distances for individual
galaxies prove invaluable in mapping out the cluster structure.
Because of its prime importance in establishing the
extragalactic distance scale, searches for Cepheids were carried out successfully in several
spiral galaxies belonging to the Virgo Cluster, using both ground-based telescopes and the post-refurbishment

Unfortunately, with typical errors of $\approx$ 0.3~mag (TF) and 0.45~mag (FP)~mag, 
the Fundamental Plane and Tully-Fisher distance indicators do not have
the accuracy needed to resolve structure in the inner regions of the cluster (see Yasuda \etal 1997;
Gavazzi \etal 1999). For instance, if the cluster is approximately spherical in shape,
then the $rms$ scatter of early-type member galaxies on the plane of the sky translates
into an $rms$ depth along the line of sight of $\sim$ 0.1~mag. Thus, to fully resolve the 
structure of the cluster --- including the core region which is dominated by early-type
dwarf galaxies (Binggeli, Tammann \& Sandage 1987) --- distances with an accuracy of
$\lesssim 0.1$~mag are required. The method of surface brightness fluctuations
(SBF) offers an attractive route forward, since it is possible to
measure distances of this accuracy for a large number of early-type galaxies in 
a relatively modest allocation of observing time.

The method itself was devised by Tonry \& Schneider (1988) and is based on the fact
that the Poissonian distribution of unresolved stars in a galaxy produces fluctuations
in each pixel of the galaxy image. Since its introduction, the method has been employed
by many groups to measure distances for early-type galaxies in the Virgo Cluster,
using data from both ground--based telescopes (Tonry et al.\ 1990, 2000, 2001;
Pahre \& Mould 1994; Jensen et al.\ 1998; Jerjen et al. 2004) and from \hst\
(Ajhar et al.\ 1997; Neilsen \& Tsvetanov 2000; Jensen et al.\ 2003). These
data have yielded some important insights into the three-dimensional structure
of the cluster.

West \& Blakeslee (2000) used the SBF catalog of Tonry \etal (2001)
to investigate the cluster's {\it principal axis} and its relation to
the surrounding large--scale structure. They found that the brightest
ellipticals lie on an axis, inclined $\sim 10^{\circ}$--$15^{\circ}$ from
the line of sight.
More recently, Jerjen \etal (2004) measured SBF distances for 16
Virgo dwarf galaxies, calibrated with stellar population model predictions,
and found distances ranging between 14.9 and 21.3~Mpc. They identified two 
clumps of galaxies, associated with M87 and M86, and determined a
back-to-front cluster depth of 6~Mpc (i.e., 2$\sigma$ of their galaxy
distance moduli distribution).
This work effectively ruled out an earlier conclusion
by Young \& Currie (1995) that the depth of the Virgo early-type dwarfs
was about twice as large.
Our recent distance measurements based on the half--light radii of globular
clusters also argue against such an extremely elongated distribution
(see Jord\'an \etal 2005a).

In this paper, we present SBF distance measurements for galaxies in the
ACS Virgo Cluster Survey (ACSVCS; Cot\'e \etal 2004; Paper~I). This large {\it HST}
program was designed, in part, to yield homogeneous SBF distances of the
highest possible precision for a large sample of early-type galaxies in the Virgo Cluster. 
We combine our SBF distances with updated radial velocity measurements from the 
literature to examine the depth and three-dimensional structure of the cluster.
Previous papers in this series have discussed the
data reduction pipeline (Jord\'an \etal 2004a = Paper II),
the connection between low-mass X-ray binaries in M87 (Jord\'an \etal
2004b = Paper III), the measurement and calibration of surface
brightness fluctuation magnitudes (Mei \etal 2005ab = Papers IV and V),
the morphology, isophotal parameters and surface brightness
profiles for early-type galaxies (Ferrarese \etal 2006a = Paper VI),
the connection between globular clusters and ultra-compact dwarf
galaxies (Ha\c{s}egan \etal 2005 = Paper VII), the nuclei of early-type 
galaxies (C\^ot\'e \etal 2006 = Paper VIII) the color distributions
of globular clusters (Peng \etal 2006a = Paper IX), the half light radii
of globular clusters and their use as a distance indicator (Jord\'an \etal
2005a = Paper X), diffuse star clusters in early-type
galaxies (Peng \etal 2006b = Paper XI) and the connection between
supermassive black holes and central stellar nuclei in early-type
galaxies (Ferrarese \etal 2006b), and the luminosity functions and 
color--magnitude relations for globular clusters in early-type
galaxies (Jordan et al. 2006a; Jordan et al. 2006b = Paper XII; Mieske et al. 2006b = Paper XIV).

Throughout this paper, we adopt the standard $\Lambda$CDM cosmology ---
$\Omega_m h^2 = 0.127^{+0.007}_{-0.013}$ and $h = 0.73\pm0.03$ ---
from Spergel et al. (2006).

\begin{figure}
\plotone{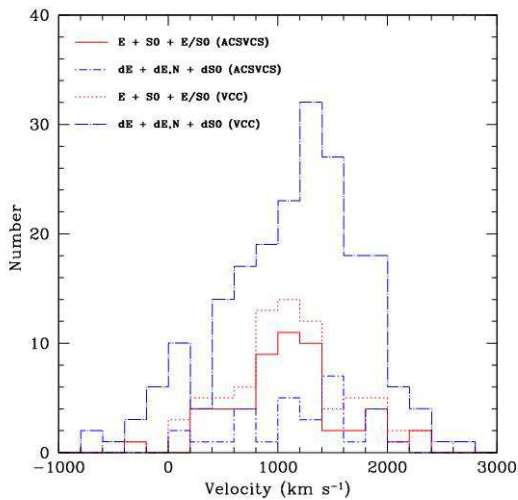}
\caption{Velocity distributions for giant and dwarf galaxies from the Virgo
Cluster Catalog (VCC) compared with those from the ACS Virgo Cluster Survey (ACSVCS). 
\label{vel}}
\end{figure}

{\it Hubble Space Telescope} ({\it HST}; Pierce \etal 1994; Freedman \etal 1994;Saha \etal 1994; Ferrarese \etal 1996; Saha \etal 1996;
Saha \etal 1997; Gibson  \etal 1999;  Graham \etal 1999;
Macri \etal 1999), while other studies employed a variety of distance indicators to explore
the distribution of galaxies within this complex structure. Pierce \& Tully (1988) measured
Tully-Fisher distances and found evidence for an infall of {\it Cluster~B} onto {\it Cluster A},
as well as for a significant depth along the line of sight. A significant depth was also
suggested by the Tully-Fisher observations of Fukugita et al. (1993). Yasuda et al. (1997)
and Federspiel et al. (1998) determined that the W and M clouds are more distant than the 
bulk of the Virgo galaxies.
Gavazzi et al. (1999), using a sample of Fundamental Plane distances for 59 early-type galaxies
and Tully-Fisher distances for 75 late-type galaxies, presented an updated cloud denomination. 
They argued that {\it Cluster~B} is infalling on {\it Cluster A} at about 750~\kms  at a relative
distance of $1.0\pm0.1$~mag, and placed the W and M clouds at $\sim$ twice the distance of
{\it Cluster~A}. They also proposed a new subclassification for {\it Cluster~A}, dividing this
structure into four regions: A, corresponding to M87 and the original {\it Cluster~A}; 
North--West (N); East (E); and South (S). Fouqu\'e \etal (2001) and Solanes \etal (2002)
studied the distribution of HI deficient galaxies in these regions.

\begin{figure*}
\epsscale{1.2}
\centerline{\plottwo{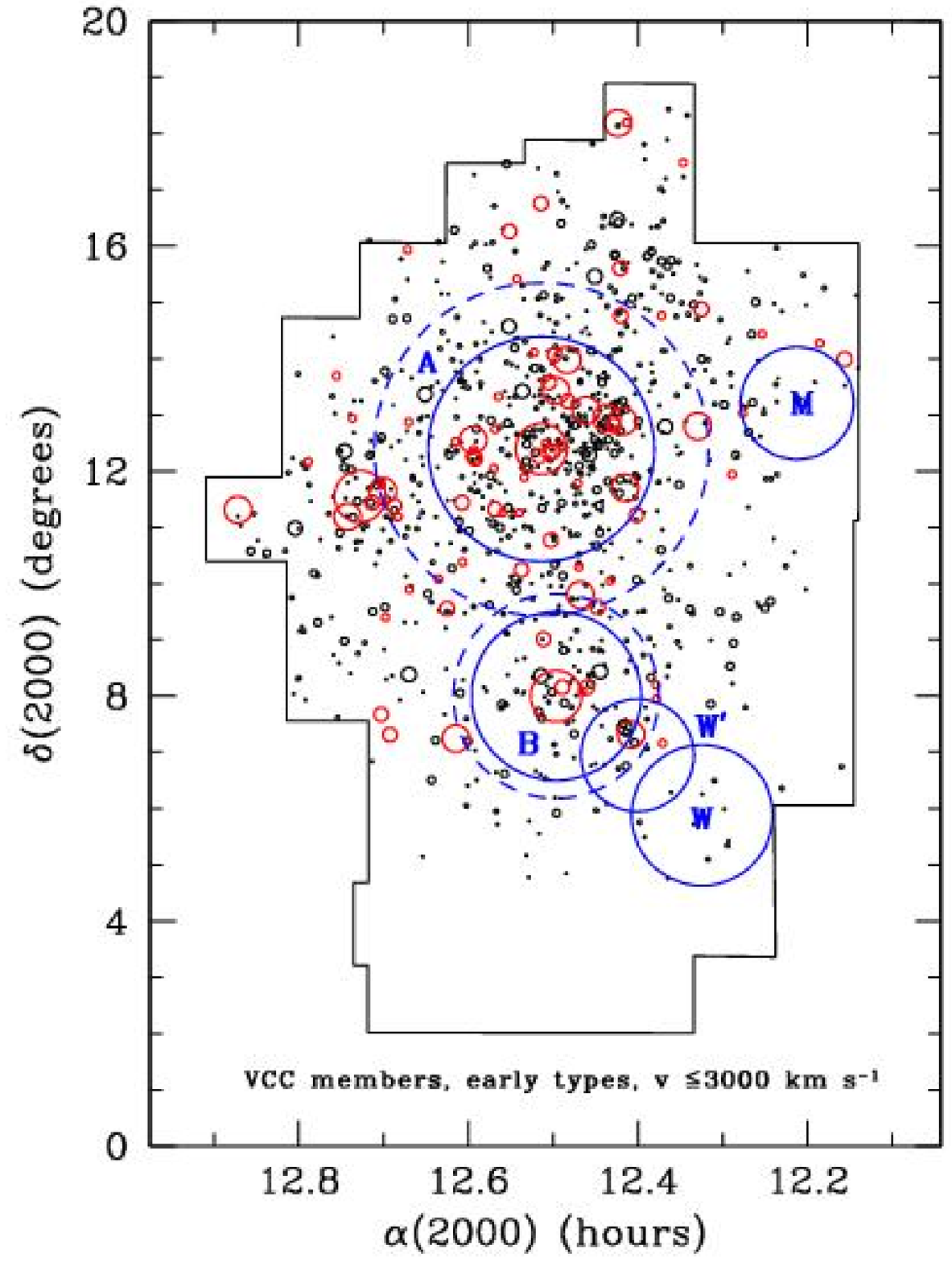}{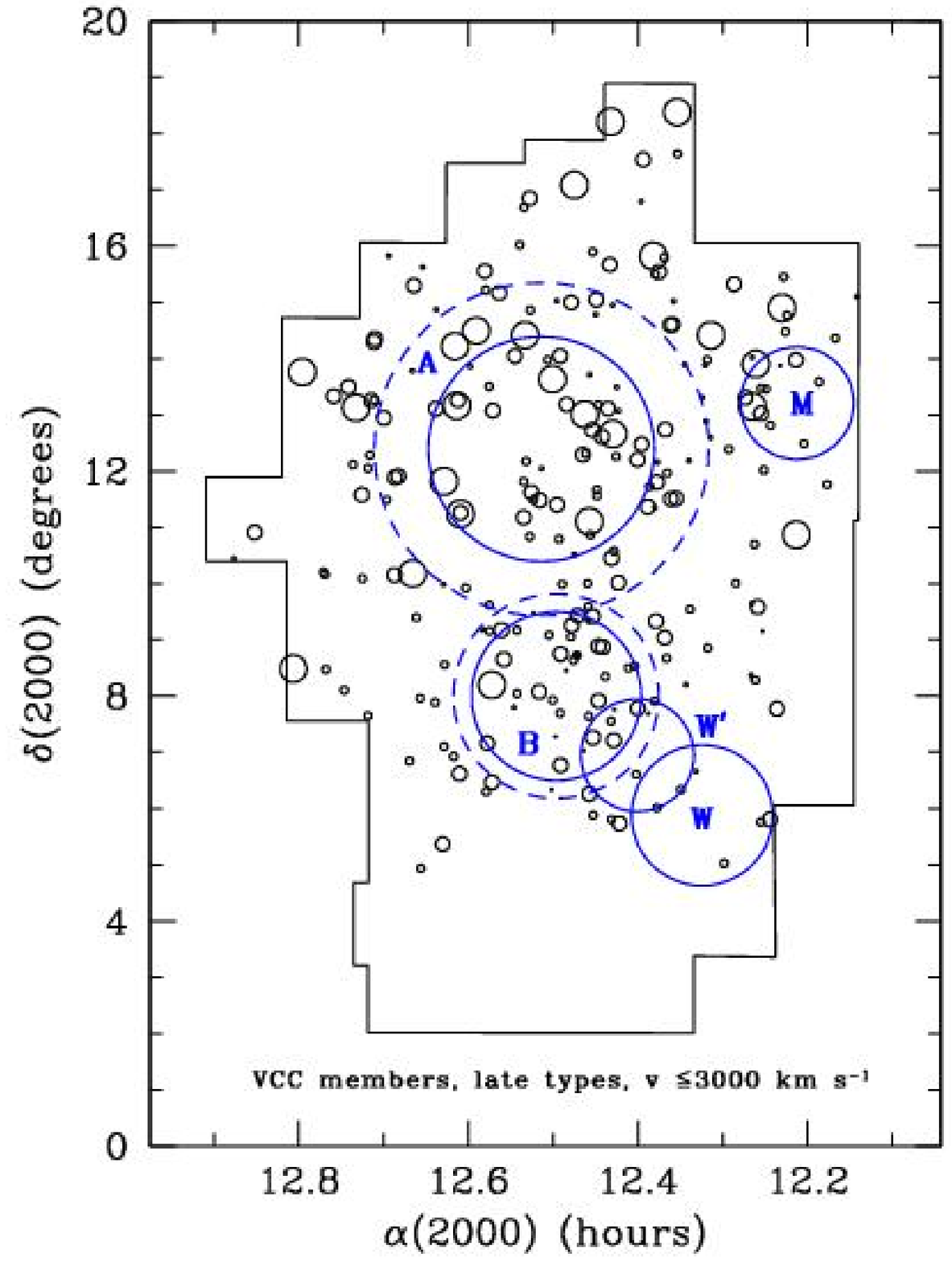}}
\caption{The distribution on the sky of {\it member} galaxies from the Virgo Cluster
Catalog (VCC) divided on the basis of morphology.
{\it (Left Panel)} 938 galaxies with {\it early-type} morphologies (i.e., E, S0, dE, dE,N,
dS0, or dS0,N) and having $v_r \le 3000$ \kms. 
Red symbols show the 100 early-type galaxies from the ACS Virgo Cluster Survey.
{\it (Right Panel)} Same as the previous panel, except for 225 member galaxies
with $v_r \le 3000$ \kms~and {\it late-type} morphologies.
\label{fig3}}
\end{figure*}

\section{The Selection of ACSVCS Program Galaxies}

Because of its richness 
and proximity, the Virgo Cluster covers an area of more than 100 deg$^2$ on the sky. 
It has a complex and irregular structure, with galaxies of different morphological
type showing different spatial and kinematic distributions (e.g., de Vaucouleurs 1961;
Sandage \& Tammann 1976; Huchra 1985; Binggeli \etal 1987, 1993; Yasuda \etal 1997;
Gavazzi \etal 1999; Solanes \etal 2002; Boselli \& Gavazzi 2006). Moreover, there is
evidence for systematic differences between the spatial distribution and kinematics
of early-type dwarf and giant galaxies in the cluster (Binggeli \etal 1987, 1993;
Binggeli 1999). As a further complication, the cluster is projected against a number
of distinct foreground and background structures, making it difficult to 
disentangle its structure from that of the surrounding large-scale environment.
In this section, we briefly review the global structure of the cluster and its
relationship to the adjacent clouds, and describe the selection function of the ACSVCS galaxies 
that we use to examine the three-dimensional structure of the cluster.

Figure~\ref{fig1} shows the distribution on the sky of galaxies classified as
{\it members} (left panel) and {\it possible members} (right panel) of the Virgo
Cluster, using classifications from the Virgo Cluster Catalog of Binggeli \etal (1985)
and excluding galaxies with heliocentric radial velocities\footnote{Based on 
radial velocity measurements for VCC galaxies compiled from the NASA Extragalactic
Database as of May 2006, and including new spectroscopic measurements from the
fourth and earlier data releases of the Sloan Digital Sky Survey.}
of $v_r \ge 3000$ \kms.

In both panels, the symbol size is proportional to blue luminosity. The
continuous blue circles identify the $A$ and $B$ Clusters and $M$, $W$ and 
$W^{\prime}$ Clouds as defined in Binggeli \etal (1987). For comparison, the
broken blue circles show the $r_{200}/2$ for Clusters A and B, as determined
from modeling of X-ray mass profiles of M87/Cluster A and M49/Cluster B from
ROSAT, optical surface brightness and integrated-light velocity dispersion
profiles for M49 and M87 (McLaughlin 1999; C\^ot\'e \etal 2001, 2003). These
two ``subclusters" are evident in the right panel of Figure~\ref{fig1}. Note
the dominance of Cluster A relative  to Cluster B (e.g., Schindler \etal
1999). The M, W and \wprime Clouds are apparent in the left panel of
Figure~\ref{fig1}. The first two of these clouds are thought to lie ``at
roughly twice the distance of the Virgo Cluster" (Binggeli \etal 1993)
while the \wprime structure is believed to connect Cluster B with
the W Cloud. Further to the south, at declinations of $\delta \lesssim 5^{\circ}$,
the Southern Extension (SE) of Virgo defines a filamentary ``spur" that
may extend towards the background (e.g., Tully 1982; Hoffman \etal 1995).

The red circles in the left panel of Figure~\ref{fig1} show the 100 early-type
galaxies from the ACSVCS. As described in Paper~I, these galaxies were drawn
from the sample of 163 VCC member galaxies (confirmed by radial velocity measurements)
which have blue magnitudes brighter than $B_T = 16$ and with early-type
morphological classifications (E, S0, dE, dE,N, dS0 or dS0,N). A total of
63 of these galaxies were excluded from the ACSVCS sample either because
they were included in previous {\it HST} (WFPC2) programs, or because of
the lack of a clearly visible
bulge component, the presence of strong dust lanes, or signs of strong
tidal interactions. The final sample of 100 galaxies includes no objects 
with declinations $\delta < 7^{\circ}$, to avoid contamination
from the Southern Extension. 
As Figure~\ref{fig1} shows, most members of
the M, W and \wprime Clouds are classified as {\it possible} members of Virgo,
so our sample should be relatively free from contamination by these structures.
However, denominations based on velocity or surface brightness are really
``best guesses" (Binggeli \etal 1993) and it is likely that some interlopers
from these clouds will appear in our sample. As we will show below, the SBF 
distances indicate that five galaxies originally classified as members of 
Cluster B are almost certainly associated with the \wprime Cloud.

\begin{figure*}
\epsscale{1.2}
\centerline{\plottwo{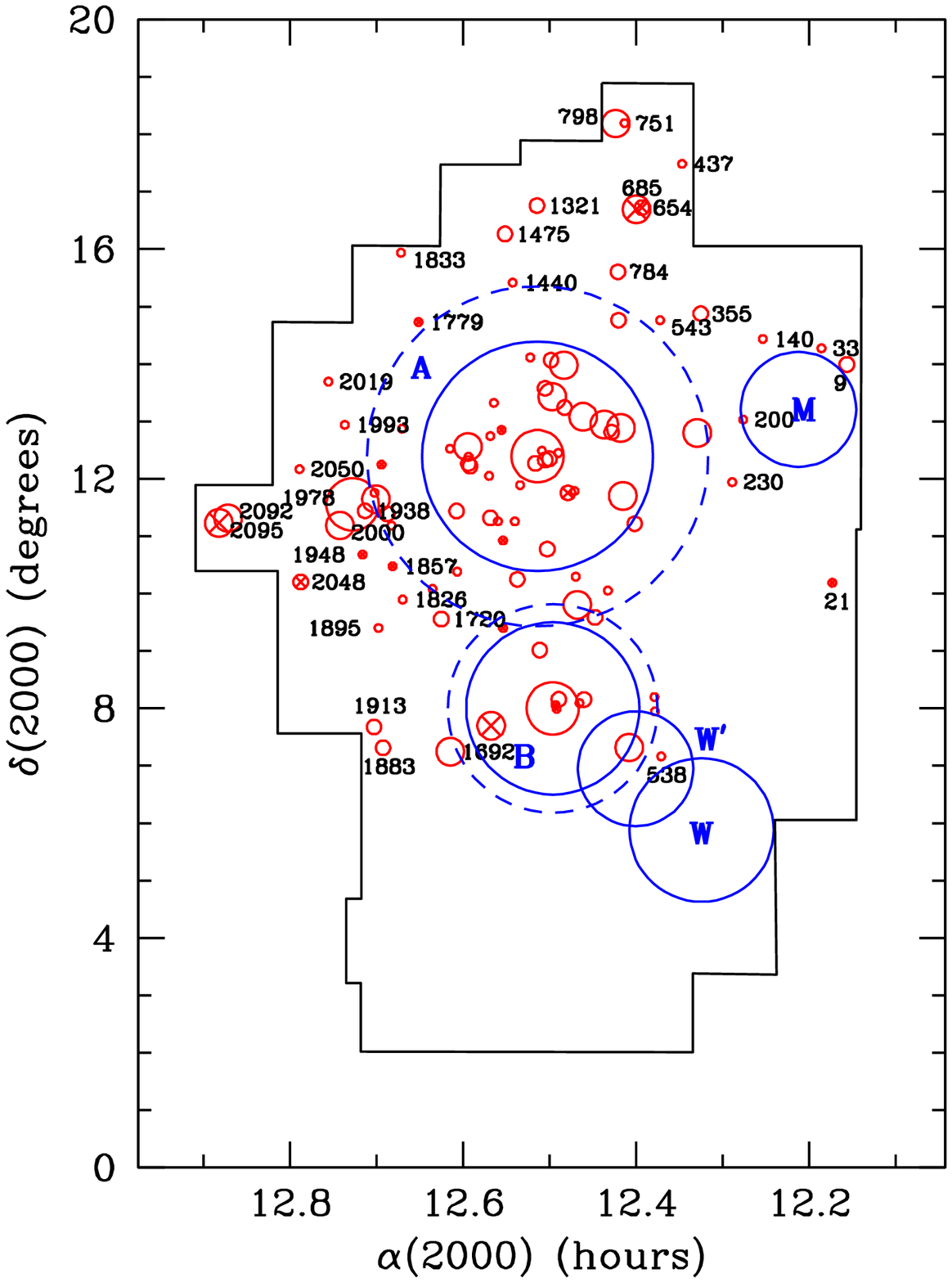}{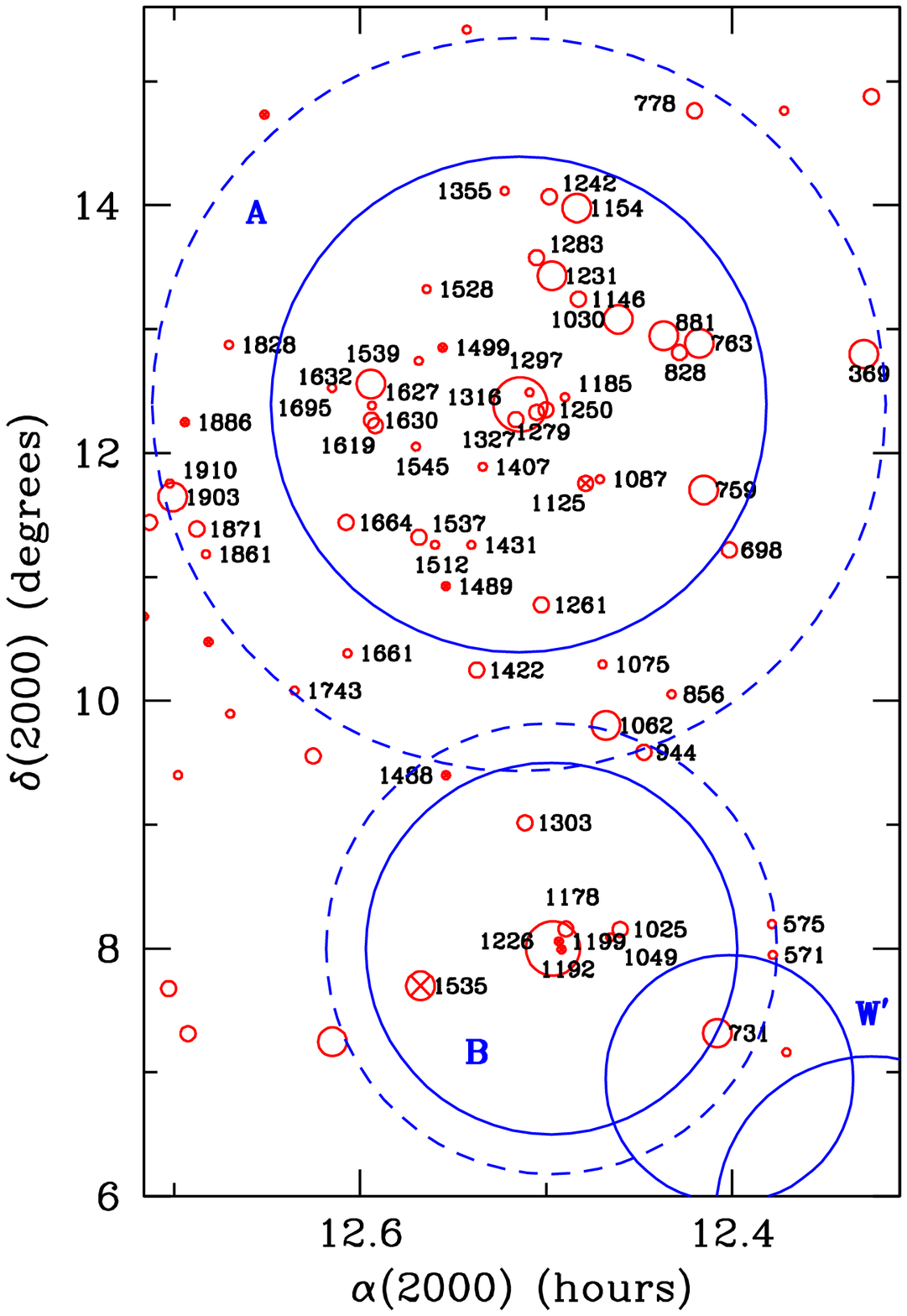}}
\caption{The distribution of galaxies from the ACS Virgo Cluster Survey.
{\it (Left Panel)} Finding charts for galaxies beyond
$r_{200}/2$ for Clusters A and B (centered on M87 and M49, respectively).
Circles with crosses show galaxies for which it was not possible to
measure a reliable SBF distance. 
{\it (Right Panel)} A magnified view of the regions centered on M87 and M49,
with the remaining galaxies from the ACS Virgo Cluster Survey labeled.
Circles with crosses show those galaxies for which it was not possible to
measure a reliable SBF distance.
\label{finder1}}
\end{figure*}

Finding charts for the 100 ACSVCS
program galaxies are presented in Figure~\ref{finder1}. As we explain below,
for 16 of the galaxies in the survey, it was not possible to derive a reliable
SBF distance. These galaxies are highlighted with crosses in Figure~\ref{finder1}.
\section{SBF Distance Measurements}

The ACSVCS data reduction and SBF measurement procedures have been fully described
in Papers~II,~IV and V. In this section, we briefly summarize the main steps involved
in the determination of SBF distances.

\subsection{SBF Magnitudes}

The SBF method was introduced by Tonry \& Schneider (1988; for reviews see
Jacoby et al. 1992 and Blakeslee et al. 1999). The SBF are defined as the
variance of the normalized Poissonian fluctuations of the galaxy stellar
population. The variance is normalized by the galaxy surface brightness,
then converted to a magnitude, called \mbar.
The measured \mbar\ depends on galaxy distance, and varies as a function
of the stellar population age and metallicity. To obtain distance measurements,
SBF magnitudes must therefore be carefully calibrated in terms of stellar population
observables --- usually galaxy colors.  Tonry \etal (1997, 2001) have shown that
$I$-band SBF magnitudes in elliptical galaxies can be calibrated as a function
of the $(V-I)$ galaxy color over the range $1 < (V-I)_0 < 1.3$~mag, with an
internal scatter $\lta 0.1$~mag. This general approach has been used in many
programs to measure early-type galaxy distances with both ground--based
telescopes and \hst\ (Ajhar \etal\ 1997, 2001; Tonry \etal\ 1997, 2001; Jensen
\etal\ 1999, 2003; Blakeslee \etal\ 2001, 2002; Mei \etal\ 2001, 2003; Liu
\etal\ 2001, 2002; Mieske \etal 2003; Cantiello \etal\ 2005;
and references in Paper~IV).

Each of the 100 ACSVCS galaxies was observed for a single orbit with {\it HST}, for a
total of 750~sec in F475 ($g_{475}$) and 1210~sec in F850LP ($z_{850}$).
The choice of the $g_{475}$ and $z_{850}$ bandpasses was dictated by their high throughput and
sensitivity to changes in stellar population age 
and metallicity. Our SBF measurements were performed in the $z_{850}$ filter
and calibrated in terms of the 
$(g_{475}-z_{850})_0$ galaxy color. As shown in
Figure~7 of Paper~I, which is based on model predictions from Blakelee \etal (2001), SBF
measurements in this filter will be brighter, and show less scatter, than corresponding 
measurements made in the $V$, $R$ or $I$-filters. Our data reduction procedures account 
for the significant geometrical distortions caused by the off-axis location of the ACS/WFC
in the {\it HST} focal plane and were optimized to guard against possible biases
introduced by resampling the pixel values (see Papers~II and~IV).

Radial velocity histograms for the ACSVCS sample (subdivided into giants and
dwarfs) are compared to those from the VCC in Figure~\ref{vel}. There is good
agreement between the ASCVCS and VCC samples, and we conclude
from Figures~\ref{fig1} and \ref{vel} that
the ACSVCS sample should be a reliable tracer population for the
early-type galaxies in the main body of the Virgo Cluster.
At the same time, though, Figure~\ref{fig3} serves as a clear reminder that
the cluster shows unmistakable evidence for a morphology-density relation,
with the late-type systems being much less centrally concentrated than
the early-type galaxies studied here (Binggeli et al. 1987).

\section{The Selection of ACSVCS Program Galaxies}

Because of its richness 
and proximity, the Virgo Cluster covers an area of more than 100 deg$^2$ on the sky. 
It has a complex and irregular structure, with galaxies of different morphological
type showing different spatial and kinematic distributions (e.g., de Vaucouleurs 1961;
Sandage \& Tammann 1976; Huchra 1985; Binggeli \etal 1987, 1993; Yasuda \etal 1997;
Gavazzi \etal 1999; Solanes \etal 2002; Boselli \& Gavazzi 2006). Moreover, there is
evidence for systematic differences between the spatial distribution and kinematics
of early-type dwarf and giant galaxies in the cluster (Binggeli \etal 1987, 1993;
Binggeli 1999). As a further complication, the cluster is projected against a number
of distinct foreground and background structures, making it difficult to 
disentangle its structure from that of the surrounding large-scale environment.
In this section, we briefly review the global structure of the cluster and its
relationship to the adjacent clouds, and describe the selection function of the ACSVCS galaxies 
that we use to examine the three-dimensional structure of the cluster.

Figure~\ref{fig1} shows the distribution on the sky of galaxies classified as
{\it members} (left panel) and {\it possible members} (right panel) of the Virgo
Cluster, using classifications from the Virgo Cluster Catalog of Binggeli \etal (1985)
and excluding galaxies with heliocentric radial velocities\footnote{Based on 
radial velocity measurements for VCC galaxies compiled from the NASA Extragalactic
Database as of May 2006, and including new spectroscopic measurements from the
fourth and earlier data releases of the Sloan Digital Sky Survey.}
of $v_r \ge 3000$ \kms. 
In both panels, the symbol size is proportional to blue luminosity. The
continuous blue circles identify the $A$ and $B$ Clusters and $M$, $W$ and 
$W^{\prime}$ Clouds as defined in Binggeli \etal (1987). For comparison, the
broken blue circles show the $r_{200}/2$ for Clusters A and B, as determined
from modeling of X-ray mass profiles of M87/Cluster A and M49/Cluster B from
ROSAT, optical surface brightness and integrated-light velocity dispersion
profiles for M49 and M87 (McLaughlin 1999; C\^ot\'e \etal 2001, 2003). These
two ``subclusters" are evident in the right panel of Figure~\ref{fig1}. Note
the dominance of Cluster A relative  to Cluster B (e.g., Schindler \etal
1999). The M, W and \wprime Clouds are apparent in the left panel of
Figure~\ref{fig1}. The first two of these clouds are thought to lie ``at
roughly twice the distance of the Virgo Cluster" (Binggeli \etal 1993)
while the \wprime structure is believed to connect Cluster B with
the W Cloud. Further to the south, at declinations of $\delta \lesssim 5^{\circ}$,
the Southern Extension (SE) of Virgo defines a filamentary ``spur" that
may extend towards the background (e.g., Tully 1982; Hoffman \etal 1995).

The red circles in the left panel of Figure~\ref{fig1} show the 100 early-type
galaxies from the ACSVCS. As described in Paper~I, these galaxies were drawn
from the sample of 163 VCC member galaxies (confirmed by radial velocity measurements)
which have blue magnitudes brighter than $B_T = 16$ and with early-type
morphological classifications (E, S0, dE, dE,N, dS0 or dS0,N). A total of
63 of these galaxies were excluded from the ACSVCS sample either because
they were included in previous {\it HST} (WFPC2) programs, or because of
the lack of a clearly visible
bulge component, the presence of strong dust lanes, or signs of strong
tidal interactions. The final sample of 100 galaxies includes no objects 
with declinations $\delta < 7^{\circ}$, to avoid contamination
from the Southern Extension. 
As Figure~\ref{fig1} shows, most members of
the M, W and \wprime Clouds are classified as {\it possible} members of Virgo,
so our sample should be relatively free from contamination by these structures.
However, denominations based on velocity or surface brightness are really
``best guesses" (Binggeli \etal 1993) and it is likely that some interlopers
from these clouds will appear in our sample. As we will show below, the SBF 
distances indicate that five galaxies originally classified as members of 
Cluster B are almost certainly associated with the \wprime Cloud.

Radial velocity histograms for the ACSVCS sample (subdivided into giants and
dwarfs) are compared to those from the VCC in Figure~\ref{vel}. There is good
agreement between the ASCVCS and VCC samples, and we conclude
from Figures~\ref{fig1} and \ref{vel} that
the ACSVCS sample should be a reliable tracer population for the
early-type galaxies in the main body of the Virgo Cluster.
At the same time, though, Figure~\ref{fig3} serves as a clear reminder that
the cluster shows unmistakable evidence for a morphology-density relation,
with the late-type systems being much less centrally concentrated than
the early-type galaxies studied here (Binggeli et al. 1987). 

Finding charts for the 100 ACSVCS
program galaxies are presented in Figure~\ref{finder1}. As we explain below,
for 16 of the galaxies in the survey, it was not possible to derive a reliable
SBF distance. These galaxies are highlighted with crosses in Figure~\ref{finder1}.

\begin{figure}
\epsscale{1.3}
\plotone{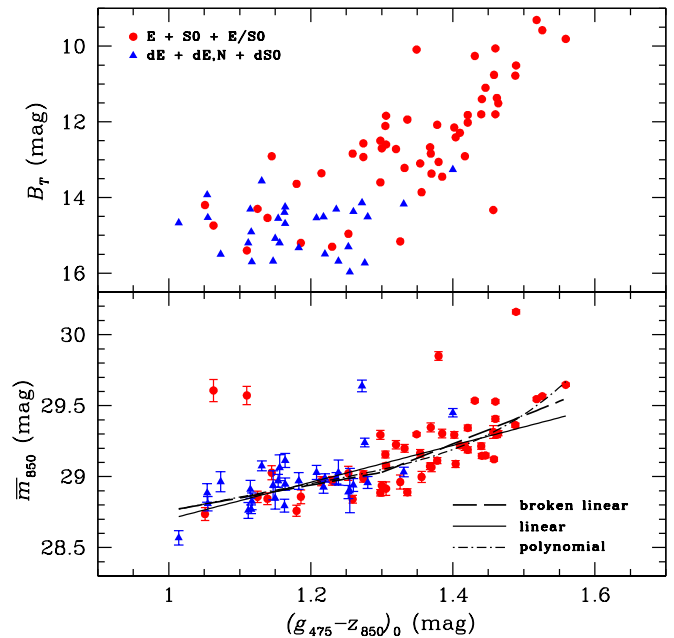}
\caption{{\it (Upper Panel)} Color magnitude diagram in the $B_T$-$(g_{\rm 475}-z_{\rm 850})_0$ plane for
84 galaxies from the ACS Virgo Cluster Survey with measured SBF magnitudes (see Table~1). Galaxies with
giant and dwarf classifications from Binggeli \etal (1985) are plotted as red circles and blue triangles,
respectively.
{\it (Lower Panel)} SBF magnitude, $\overline{m}_{\rm 850}$, as a function of galaxy color for the 84
galaxies shown in the upper panel. Three different SBF measurement calibrations are overlaid: (1) a
broken linear relation matched at a color
of $(g_{\rm 475}-z_{\rm 850})_0 = 1.3$~mag (heavy dashed line; Equation~\ref{eq:broken}) (2) a single
linear relation (solid line; Equation~\ref{eq:linear}); and (3) a fourth-order
polynomial fit (dashed-dotted curve; Equation~\ref{eq:poly}). Galaxies with
giant and dwarf classifications from Binggeli \etal (1985) are plotted as red circles and blue triangles,
respectively. Five galaxies belonging to the W$^{\prime}$ Cloud lie $\sim$~0.7~mag above the fitted relations.
\label{calib1}}
\end{figure}

\section{SBF Distance Measurements}

The ACSVCS data reduction and SBF measurement procedures have been fully described
in Papers~II,~IV and V. In this section, we briefly summarize the main steps involved
in the determination of SBF distances.

\subsection{SBF Magnitudes}

The SBF method was introduced by Tonry \& Schneider (1988; for reviews see
Jacoby et al. 1992 and Blakeslee et al. 1999). The SBF are defined as the
variance of the normalized Poissonian fluctuations of the galaxy stellar
population. The variance is normalized by the galaxy surface brightness,
then converted to a magnitude, called \mbar.
The measured \mbar\ depends on galaxy distance, and varies as a function
of the stellar population age and metallicity. To obtain distance measurements,
SBF magnitudes must therefore be carefully calibrated in terms of stellar population
observables --- usually galaxy colors.  Tonry \etal (1997, 2001) have shown that
$I$-band SBF magnitudes in elliptical galaxies can be calibrated as a function
of the $(V-I)$ galaxy color over the range $1 < (V-I)_0 < 1.3$~mag, with an
internal scatter $\lta 0.1$~mag. This general approach has been used in many
programs to measure early-type galaxy distances with both ground--based
telescopes and \hst\ (Ajhar \etal\ 1997, 2001; Tonry \etal\ 1997, 2001; Jensen
\etal\ 1999, 2003; Blakeslee \etal\ 2001, 2002; Mei \etal\ 2001, 2003; Liu
\etal\ 2001, 2002; Mieske \etal 2003; Cantiello \etal\ 2005;
and references in Paper~IV).

Each of the 100 ACSVCS galaxies was observed for a single orbit with {\it HST}, for a
total of 750~sec in F475 ($g_{475}$) and 1210~sec in F850LP ($z_{850}$).
The choice of the $g_{475}$ and $z_{850}$ bandpasses was dictated by their high throughput and
sensitivity to changes in stellar population age 
and metallicity. Our SBF measurements were performed in the $z_{850}$ filter
and calibrated in terms of the 
$(g_{475}-z_{850})_0$ galaxy color. As shown in
Figure~7 of Paper~I, which is based on model predictions from Blakelee \etal (2001), SBF
measurements in this filter will be brighter, and show less scatter, than corresponding 
measurements made in the $V$, $R$ or $I$-filters. Our data reduction procedures account 
for the significant geometrical distortions caused by the off-axis location of the ACS/WFC
in the {\it HST} focal plane and were optimized to guard against possible biases
introduced by resampling the pixel values (see Papers~II and~IV).

\begin{figure}
\centerline{
\includegraphics[angle=90,scale=0.4]{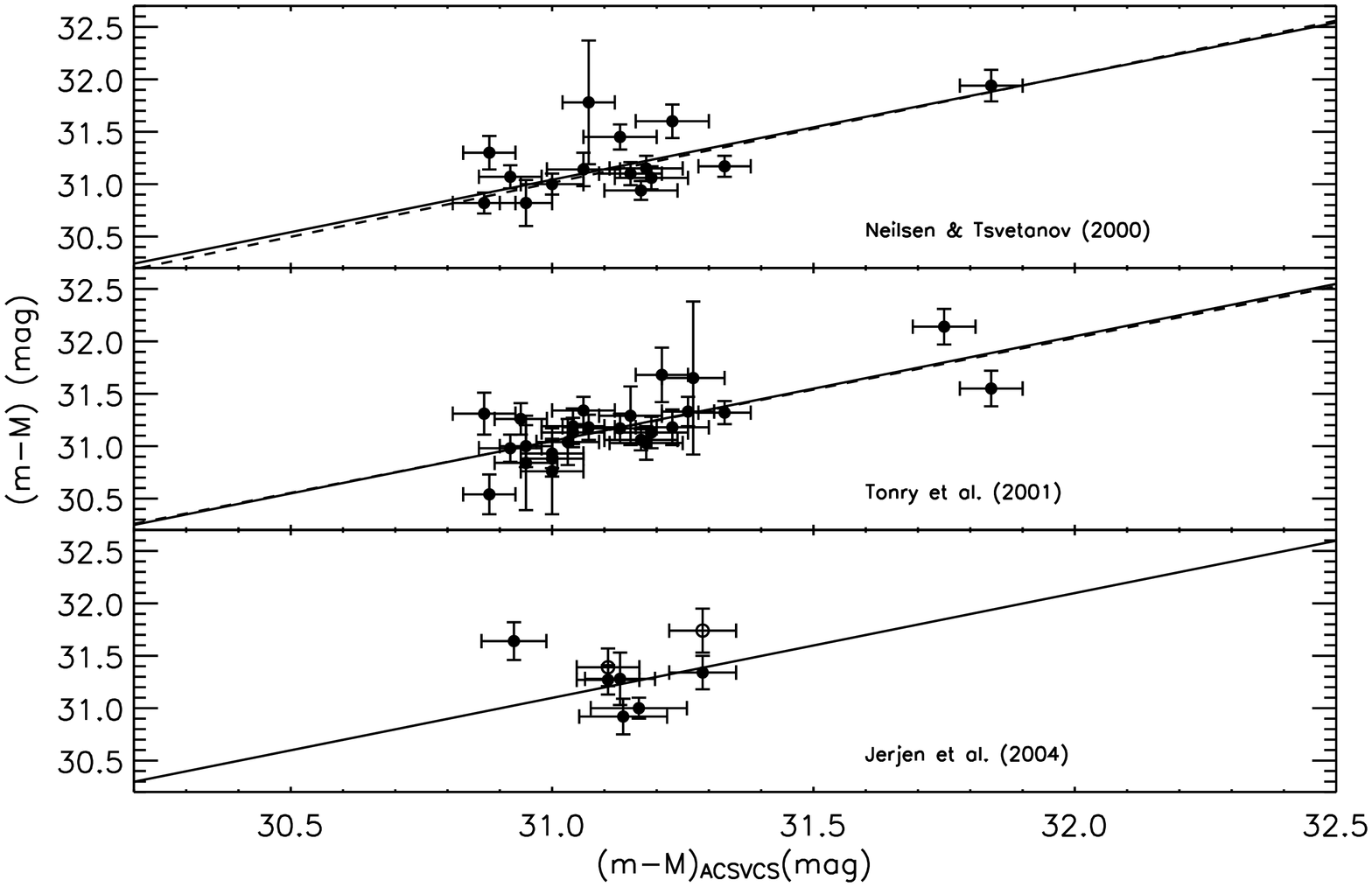}}
\caption{{\it (Upper Panel)} Comparison of distance moduli from Neilsen \& Tsvetanov  (2000) with those
from the ACS Virgo Cluster Survey for 15 Virgo galaxies in common between the two surveys.  The solid
line is the one-to-one relation, while the dashed line shows the line of best fit.
{\it (Middle Panel)} Comparison of SBF distance moduli from Tonry \etal (2001) with those
from the ACS Virgo Cluster Survey for 26 Virgo galaxies in common between the two surveys. 
The solid line is the one-to-one relation, while the dashed line shows the line of best fit.
{\it (Lower Panel)} Comparison of SBF distance moduli from Jerjen \etal (2004) with those
from the ACS Virgo Cluster Survey for eight Virgo galaxies in common between the two surveys.
For two galaxies (VCC1087 = IC3381 and VCC1261 = NGC4482), Jerjen \etal (2004)
list two values for the distance modulus. In those cases, the higher measurements are shown by open 
circles. The dashed line has been calculated using the values
from Jerjen \etal (2004) which agree best with our measurements (see Table~\ref{compare}). Note
that the abscissa is stretched relative to the ordinate, and that the ACSVCS uncertainties
are between one half and one tenth those of the ground-based measurements.
\label{comparefig}}
\end{figure}
\begin{figure*}
\epsscale{0.8}
\plotone{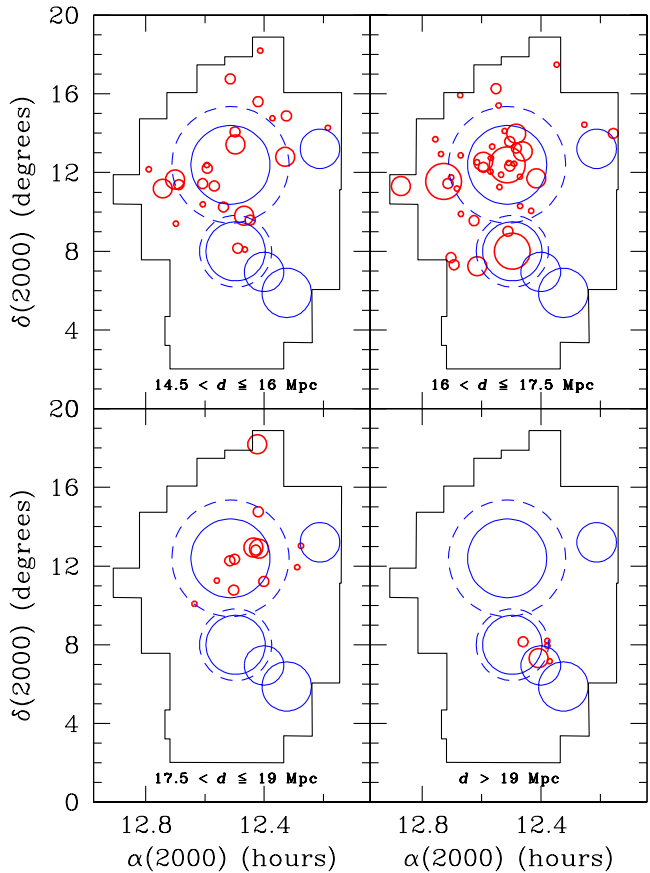}
\caption{Distribution on the sky of 84 galaxies from the ACS Virgo Cluster Survey (red
circles) with measured SBF distances,
displayed in four ranges in distance: $14.5 < d \le 16$ Mpc (upper left);
$16 < d \le 17.5$ Mpc (upper right);
$17.5 < d \le 19$ Mpc (lower left);
and $d > 19$ Mpc (lower right).
\label{distcut}}
\end{figure*}

\subsection{Choice of Distance Calibration}
\label{sec:cal}

Details on the ACSVCS SBF calibration are given in Paper~V, which presents
a broken linear calibration for the absolute SBF magnitude $\overline M_{850}$:
\begin{equation}
\begin{array}{rrrrr}
\overline M_{850}  =  -2.06 \pm 0.04& + (0.9 \pm 0.2)[ (g_{475}-z_{850})_0-1.3 ] && & \\
 &1.0 \le (g_{475}-z_{850})_0 \le 1.3; &&& \\
&&&& \\
\overline M_{850}  =  -2.06 \pm 0.04& + (2.0 \pm 0.2)[ (g_{475}-z_{850})_0-1.3 ] & & &\\
&1.3 <(g_{475}-z_{850})_0 \le 1.6 &&&\\
\end{array}
\label{eq:broken}
\end{equation}
These color regimes correspond roughly to different morphological types, with early-type giants
at the red end, and early-type dwarfs at the blue. This division is evident in the color--magnitude
diagram shown in the top panel of Figure~\ref{calib1}. Different symbols correspond to the dwarfs 
(blue triangles) and giants (red circles) in the ACSVCS sample, based on the VCC morphologies
given in Binggeli \etal (1985). While a color of $(g_{475}-z_{850})_0 = 1.3$ serves
to roughly divide the sample into dwarfs and giants, we caution that this division is
not unique. In any case, the lower panel of Figure~\ref{calib1} plots apparent SBF magnitude
against galaxy color, showing that there are no systematic differences in the
results for red or blue galaxies. The long-dashed curve shows the calibration given
by Equation~\ref{eq:broken}. The two other curves show alternate choices for the SBF
calibration; these alternate relations are discussed in the Appendix, along with a 
discussion of the possible biases and errors that may arise from our choice of calibration.

The absolute zeropoint in Equation~\ref{eq:broken} was derived from the Tonry \etal (2001)
Virgo distance modulus, corrected by the Udalski \etal (1995) Cepheid period-luminosity
relation adopted for the $H_0$ Key Project distances (Freedman \etal 2001), which gives
$31.09 \pm 0.03$~mag (see the discussion in Paper~V). The zeropoint uncertainty includes
all sources of internal error; there is an additional systematic uncertainty of
$\sim\,$0.15~mag, due to the uncertainty of Cepheid distance measurements in the
distance scale calibration. 
Comparing to stellar population models of Bruzual \& Charlot (2003), our results 
are consistent with model predictions in the range $1.3 < (g_{475}-z_{850})_0 \le 1.6$~mag. 
In the range $0.9 \le(g_{475}-z_{850})_0 \leq 1.3$~mag, the
empirical slope is somewhat steeper than the Bruzual \& Charlot model predictions.
However, as discussed by Mieske \etal\ (2006a), other models (Blakeslee \etal\ 2001;
Cantiello \etal\ 2003) predict somewhat steeper SBF-color relations in the blue,
more in line with the empirical behavior.

Using the above calibration, we derived galaxy distances for 84 galaxies in our sample. 
Distances were not measured for seven galaxies with very blue colors of
$(g_{475}-z_{850})_0 \leq 1$. In fact, the standard deviation of SBF magnitudes
increases as the $(g_{475}-z_{850})_0$ color decreases, making it difficult to establish
a reliable calibration for these blue colors (see the discussion in Paper~V).
Nine other galaxies were not included in the sample because of difficult
sky subtraction, or because they are edge--on disk or barred galaxies,
which are especially challenging for SBF measurements.

\subsection{SBF Distance Catalog}

The final SBF measurements and distance moduli from the ACSVCS are presented in
Table~\ref{catalogue}. The columns of this table record:
{\it (1)}: ID number from Cot\'e \etal (2004);
{\it (2)}: VCC number from Binggeli \etal (1985);
{\it (3)}: Galaxy $(g_{475}-z_{850})_0$ color over the same range of annuli used
for the SBF measurements (see Paper~V);
{\it (4)}: SBF magnitude $\overline m_{850}$;
{\it (5)}: Distance modulus obtained using the broken linear calibration (Equation~\ref{eq:broken});
{\it (6)}: Blue magnitude, $B_T$, from Binggeli \etal (1985);
{\it (7)}: Heliocentric radial velocity from Huchra (1985) and Binggeli \etal (1993);
{\it (8)}: Morphological type from Binggeli \etal (1985);
{\it (9)}: Identification numbers from the Messier, NGC, IC and UGC catalogs;
{\it (10)}: Numerical code to identify those galaxies without measured SBF magnitudes
or distances, or galaxies with uncertain measurements.

As in Paper~I, the identification numbers in column (1) run from the brightest to
the faintest apparent blue magnitudes. In Appendix~A, we report distances for these same
galaxies obtained with the alternate calibrations.

\subsection{Comparison with Previous SBF Surveys}

Table~\ref{compare} and  Fig.~\ref{comparefig} compare our distances with those
reported in three previous SBF surveys of the Virgo cluster: 
Neilsen \& Tsvetanov (2000), Tonry \etal (2001) and Jerjen \etal (2004). 
Note that the Tonry \etal (2001) measurements were recalibrated to the
same zeropoint that we use in this work, and that no zero-point
corrections have been applied to either the Neilsen \& Tsvetanov
(2000) or Jensen \etal (2004) datasets.

Neilsen \& Tsvetanov (2000) measured SBF distances for 15 bright early-type
Virgo cluster galaxies using
{\it HST}/WFPC2 images in the F814W bandpass calibrated against the stellar
population models of Worthey \etal (1998 private communication). These models 
gave reasonable agreement with the ground-based calibration of Tonry \etal (1997). 
The 15 galaxies from Neilsen \& Tsvetanov (2000) are also included in the ACSVCS;
the upper panel of Figure~\ref{comparefig} compares the measured distance moduli
for these objects. The dashed line shows the best-fit linear relation, accounting
for the uncertainties in both measurements (Press \etal 1992). The slope of
this relation is $1.03\pm0.18$, with an $rms$ scatter of 0.26~mag.

The survey of Tonry \etal (2001) is the largest SBF survey currently available, 
consisting of SBF measurements for 300 galaxies out to $cz \sim$~4000~km/s.
Tonry \etal (2001) measured and calibrated $I$-band SBF as a function of 
$(V-I)$ color. A total of 26 galaxies in our sample have distance measurements
from this survey. The average uncertainty in the Tonry \etal (2001) distance
moduli is 0.20~mag for these galaxies (with a median of 0.16~mag), roughly three
times larger than the average (and median) ACSVCS uncertainty of 0.07~mag. 
The measurements are compared in the middle panel of Figure~\ref{comparefig}.
The dashed line shows the best fit linear relation, which has a slope of
$0.99\pm0.16$ and an $rms$ scatter of 0.22~mag.

The Jerjen \etal (2004) sample consists of 16 dwarf galaxies, of which six are in
common with our survey. Their work made use of deep $R$ and $B$-band imaging with
FORS1 on the Very Large Telescope (VLT). Distances were derived by calibrating
the $R$-band SBF measurements as a function of $(B-R)$ color, using stellar
population models from Worthey \etal (1994) (as described in Jerjen \etal 2001).
For two galaxies in common with our survey, Jerjen  \etal (2004)
list two values for the distance modulus, depending on their choice of
calibration.
Note that stellar population models make widely varying SBF predictions 
at the blue colors of these dwarfs (e.g., Mieske \etal\ 2006a).
Given the relatively small sample of galaxies in common between the two studies,
and the fact that all six galaxies are located close to the mean cluster distance,
we do not attempt a regression analysis. The $rms$ scatter about the one-to-one
relation shown in the lower panel of Figure~\ref{comparefig} is 0.32~mag; however, 
removing the one large outlier (VCC\,1422) reduces the scatter to 0.18~mag.

We conclude from Table~\ref{compare} and Fig.~\ref{comparefig} that there is good
agreement between our distance moduli and those from previous studies. At the
same time, our new SBF distances have a mean precision that is 3--4 times better
than the previous measurements, and our sample of 84 galaxies with measured SBF
distances represents more than a three-fold increase over any single previous
SBF survey in Virgo.

\begin{figure*}
\epsscale{0.8}
\plotone{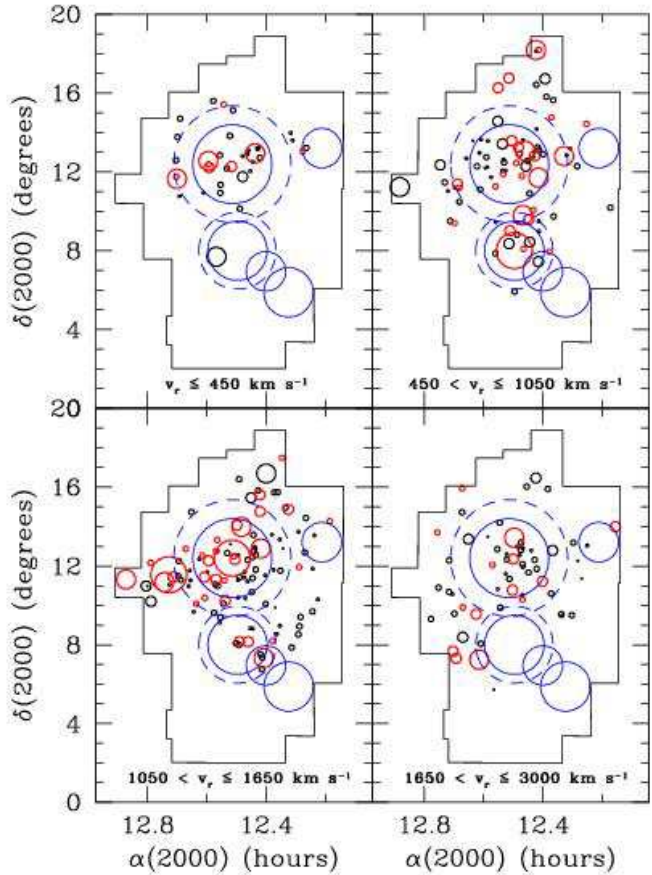}
\caption{Distribution on the sky of 84 galaxies from the ACS Virgo Cluster Survey (red
circles) with measured SBF distances,
displayed in four ranges in heliocentric radial velocity: $v_r \le 450$ \kms (upper left);
$450 < v_r \le 1050$ \kms (upper right);
$1050 < v_r \le 1650$ \kms (lower left);
and $1650 < v_r \le 3000$ \kms (lower right).
Black circles show member galaxies with early-type morphologies from the VCC.
\label{velcut}}
\end{figure*}

\section{Results}

\subsection{Cluster Structure and Line-of-Sight Depth}

Figures~\ref{distcut} and \ref{velcut} illustrate the spatial and kinematic structure 
of the cluster along the line of sight, using the SBF distances listed in Table~1. Four ``slices" in distance are shown in
Figure~\ref{distcut}: (1) $14.5 < d \le 16$ Mpc (upper left); (2) $16 < d \le 17.5$~Mpc
(upper right); (3) $17.5 < d \le 19$ Mpc (lower left); and (4) $d > 19$ Mpc. The first
of these panels shows a preponderance of galaxies in the eastern side of the cluster.
Most of the galaxies in the sample, which belong to the M87/A subcluster, appear 
in the second panel (16-17.5~Mpc). Still more distant galaxies (17.5-19~Mpc) appear in
the third panel and lie mainly in the western half of the cluster. A separate, more distant
group at $d \approx 23$~Mpc (visible in the fourth panel) is almost certainly associated
with the \wprime Cloud which is centered 
at $\alpha \approx~$12$^h$ 24$^m$ and $\delta \approx 7^{\circ}$. 
In Figure~\ref{velcut}, four ranges in heliocentric radial velocity are shown:
 $v_r \le 450$ \kms~(upper left);
$450 < v_r \le 1050$ \kms~(upper right);
$1050 < v_r \le 1650$ \kms~(lower left);
and $1650 < v_r \le 3000$ \kms~(lower right).
In the lower right panel, we notice that all the group of galaxies east (left in the figure)
of {\it Cluster~B} have $v_r  > 1650$ \kms.
Excluding the galaxies with $d \approx 23$~Mpc, Figures~\ref{distcut} and \ref{velcut} suggest that there is 
an east-to-west gradient in distance across the face of the cluster, with the eastern
portion of the cluster lying slightly in the foreground. We shall return to this issue
in \S\ref{pa}.

\begin{figure}
\epsscale{1.3}
\plotone{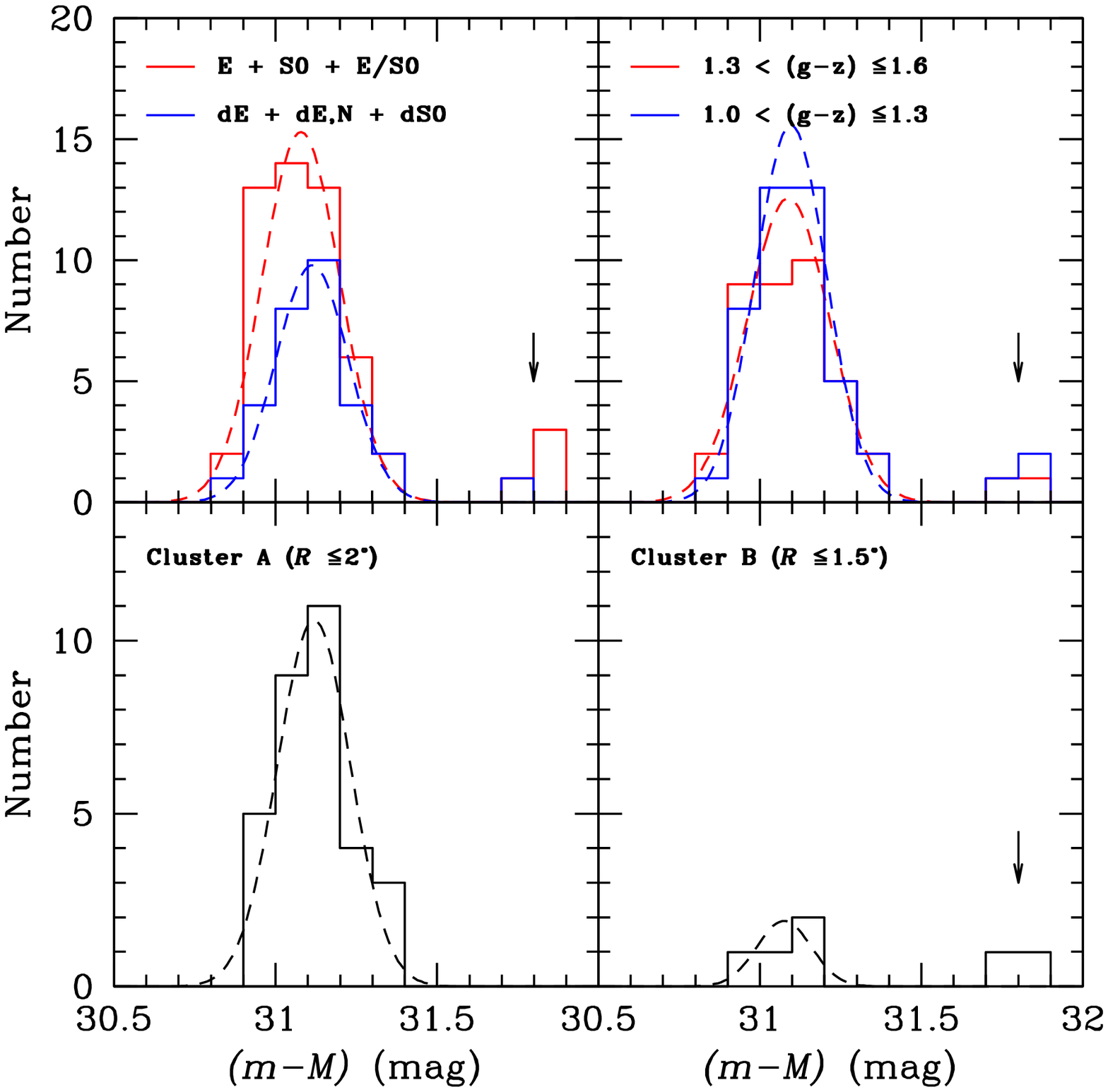}
\caption{{\it (Upper Panels)} Distribution of our measured SBF distances 
for 84 galaxies from the ACS Virgo Cluster Survey. 
The left panel shows the breakdown of the distance distributions by morphological
type, while the right panel shows the breakdown by photometric color.
The dashed curves show the best-fit Gaussian distributions.
{\it (Lower Panels)} Distribution of SBF distances for galaxies nominally
within the boundaries of ``Cluster~A'' (projected separation $<\,$2\arcmin\
from M87; left panel) and
``Cluster~B'' (projection separation $<\,$1\farcm5 from M49; right panel)
defined by Binggeli \etal (1987, 1993).
Note the presence of two galaxies (VCC1025 = NGC4434 and VCC731 = NGC4365),
that fall inside the boundaries of Cluster B but are
actually associated with the \wprime Cloud at $d \sim\,$~23~Mpc (indicated by
the arrows).
\label{substruct}}
\end{figure}

To estimate the line-of-sight depth of the cluster, we begin by dividing the ACSVCS
galaxies in various subsamples: i.e., giants vs. dwarfs, and red vs. blue galaxies.
Distance histograms for these subsamples are shown in the upper panels of
Figure~\ref{substruct}. The smooth curves in each panel show Gaussian distributions
with the parameters listed in Table~\ref{mean}. For various subsamples, the columns
of this table record the number of galaxies in each subsample, $N$, the mean distance
modulus, $\overline{(m-M)}$, the observed dispersion, $\sigma(m-M)_{\rm o}$,
the average uncertainty on the individual distance moduli, $\overline{\sigma(m-M)}_{\rm m}$, and
the {\it intrinsic} dispersion in distance modulus, $\sigma(m-M)_{\rm i}$. The
final four columns give the mean distance, $\overline{d}$, the intrinsic dispersion
in distance, $\sigma(d)$, 
mean radial velocity, $\overline{v}_r$, and the measured velocity dispersion, ${\sigma}_{v_r}$.
We take the intrinsic dispersion in distance modulus for each subsample to be
$$\sigma(m-M)_{\rm i} = \biggl[\sigma(m-M)^2_{\rm o} - \overline{\sigma(m-M)}^2_{\rm m} - \sigma(m-M)^2_{\rm c}\biggr]^{1/2}$$
where $\sigma(m-M)_{\rm c} = 0.05$~mag is the estimated ``cosmic scatter" in the fluctuation 
magnitude (e.g., Tonry \etal 1997). We see from Figure~\ref{substruct} and Table~\ref{mean}
that the \wprime Could members bias the measured depth, leading to
highly extended distributions along the line of sight. Excluding these galaxies
gives a $1\sigma$ dispersion in distance of $\approx$~0.6$\pm0.1$~Mpc, with no obvious
dependence on luminosity class or color. Our best-estimate for the $\pm2\sigma$ depth 
of the cluster is then 2.4$\pm$0.4~Mpc, a range that should include 95\% of Virgo's 
early-type galaxies. This range is about half the $\pm2\sigma$ depth of 6~Mpc
reported by Jerjen \etal (2004).

In the lower panels of Figure~\ref{substruct}, we show distance histograms for galaxies
located within the nominal boundaries of the A and B subclusters (Binggeli \etal 1987).
These two subclusters have distances that are identical, $d \approx 16.5$~Mpc, to within 
their measurement errors.

\subsection{Distance-Velocity Distribution}
\label{dv}

\begin{figure}[t!]
\epsscale{1.2}
\plottwo{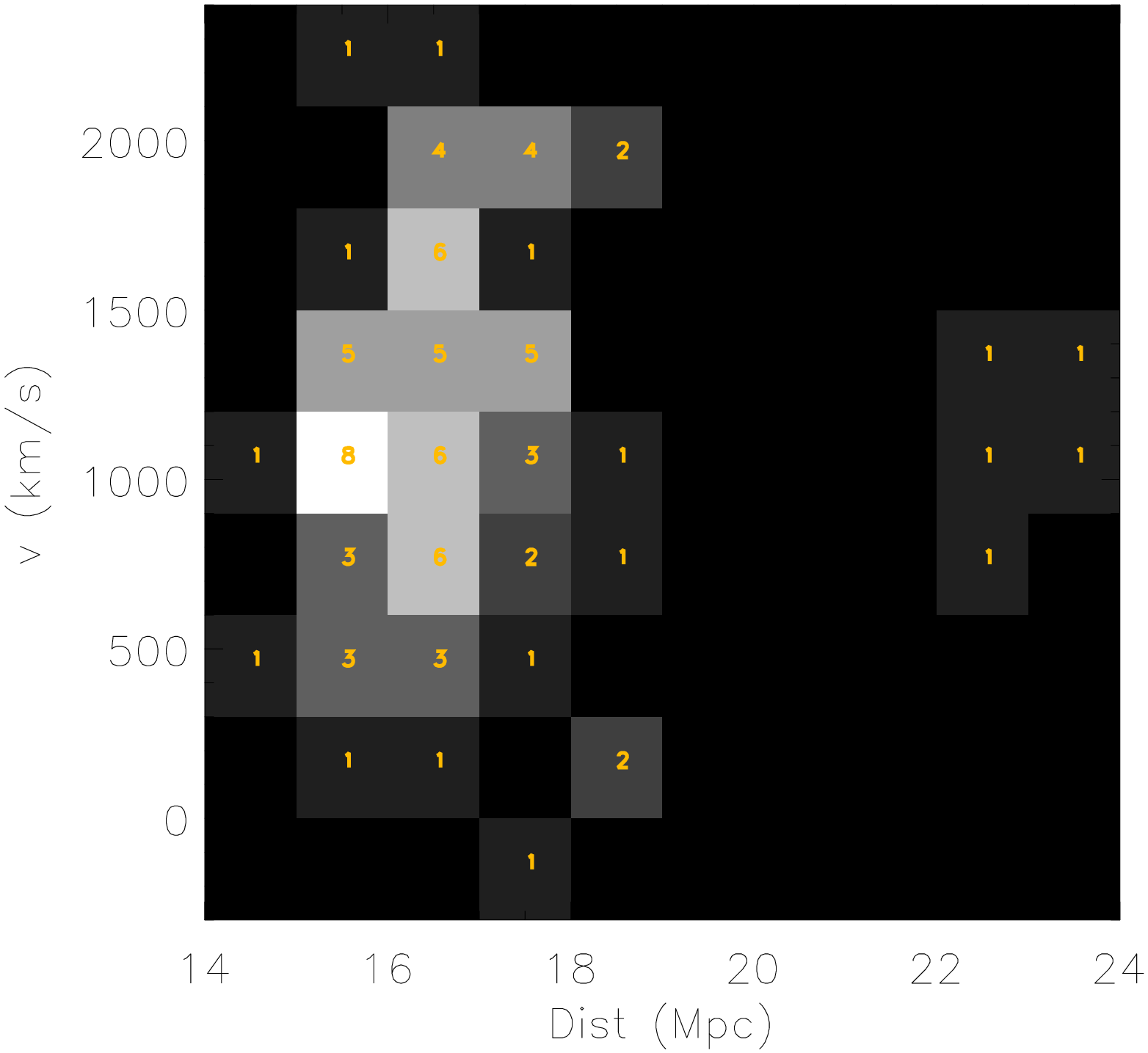}{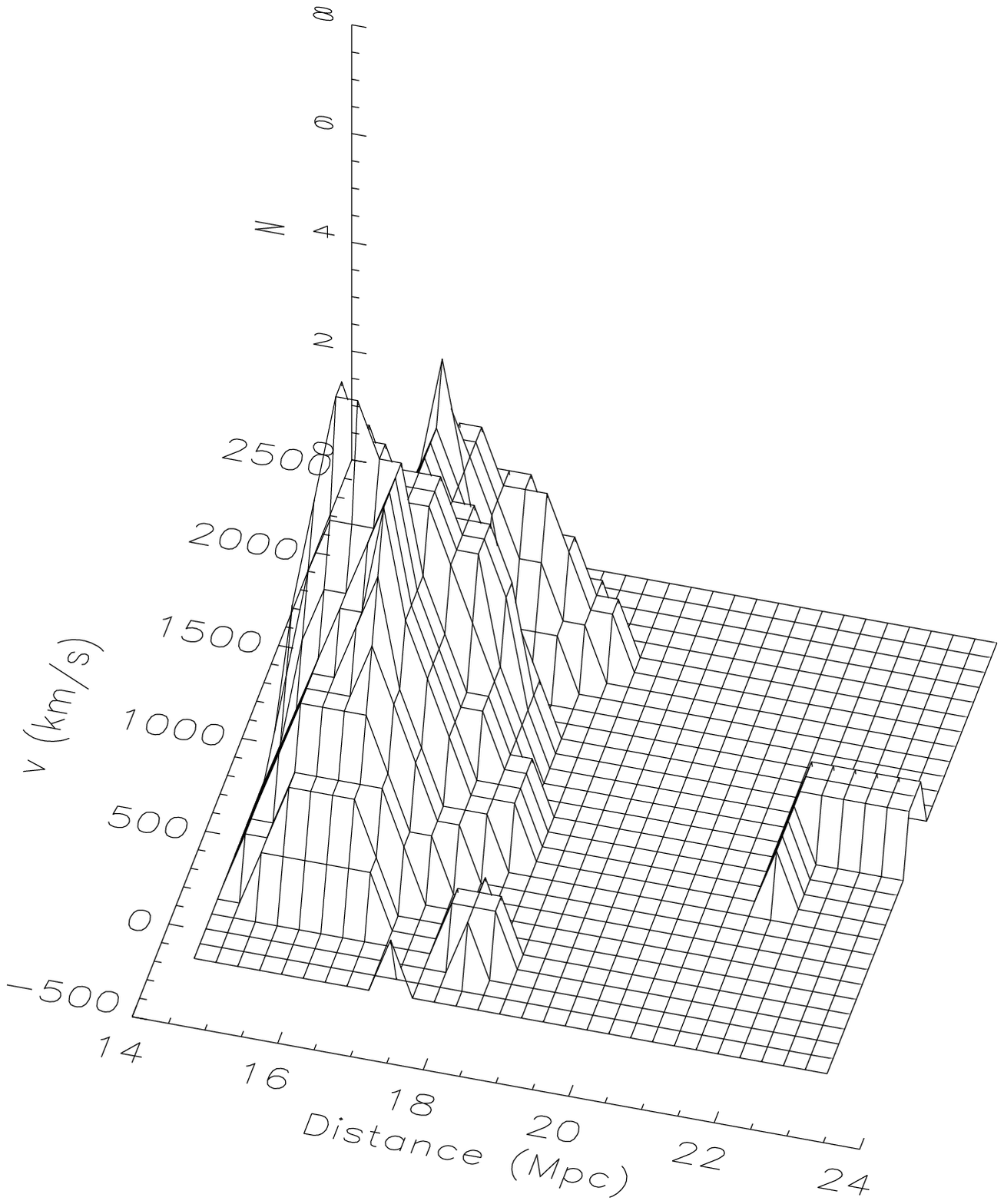}
\caption{{\it (Left Panel)} Image showing a Hubble diagram for the ACS Virgo Cluster
Survey sample, for 84 galaxies with measured SBF distances. The number of galaxies
in each ``pixel" has been labeled.
{\it (Right Panel)} Surface plot for the preceding image.
\label{group}}
\end{figure}

The distance-velocity structure of the cluster is illustrated in
Figure~\ref{group} which shows the ACSVCS galaxies, in the form of a greyscale
image (left panel) and as a surface plot (right panel).  Two obvious
substructures are apparent in this figure: (1) a large group between 15--18~Mpc
and spread over a range of $\Delta{v_r}$ $\sim$ 2000 \kms~ in radial velocity;
and (2) a small group of five galaxies at $\sim$~23~Mpc and $\sim$~1100~\kms~
which belong to the \wprime Cloud. 
A third group of galaxies associated with M86
(VCC881) may also be present at a mean distance of $17.6\pm0.6$ Mpc with
a relative velocity of about $-$1000 \kms\ with respect to the cluster mean.
However, in this case, the separation from the main
component of the cluster is less extreme, and we cannot be certain without
additional observations for an expanded sample of galaxies. We note that both
Jerjen \etal (2004) and Schindler \etal (1999) have argued that M86 represents
the dominant member of its own subcluster, albeit one much less massive than that
associated with M87 (see Table~6 of Schindler \etal 1999).

\begin{figure*}
\epsscale{0.7}
\plotone{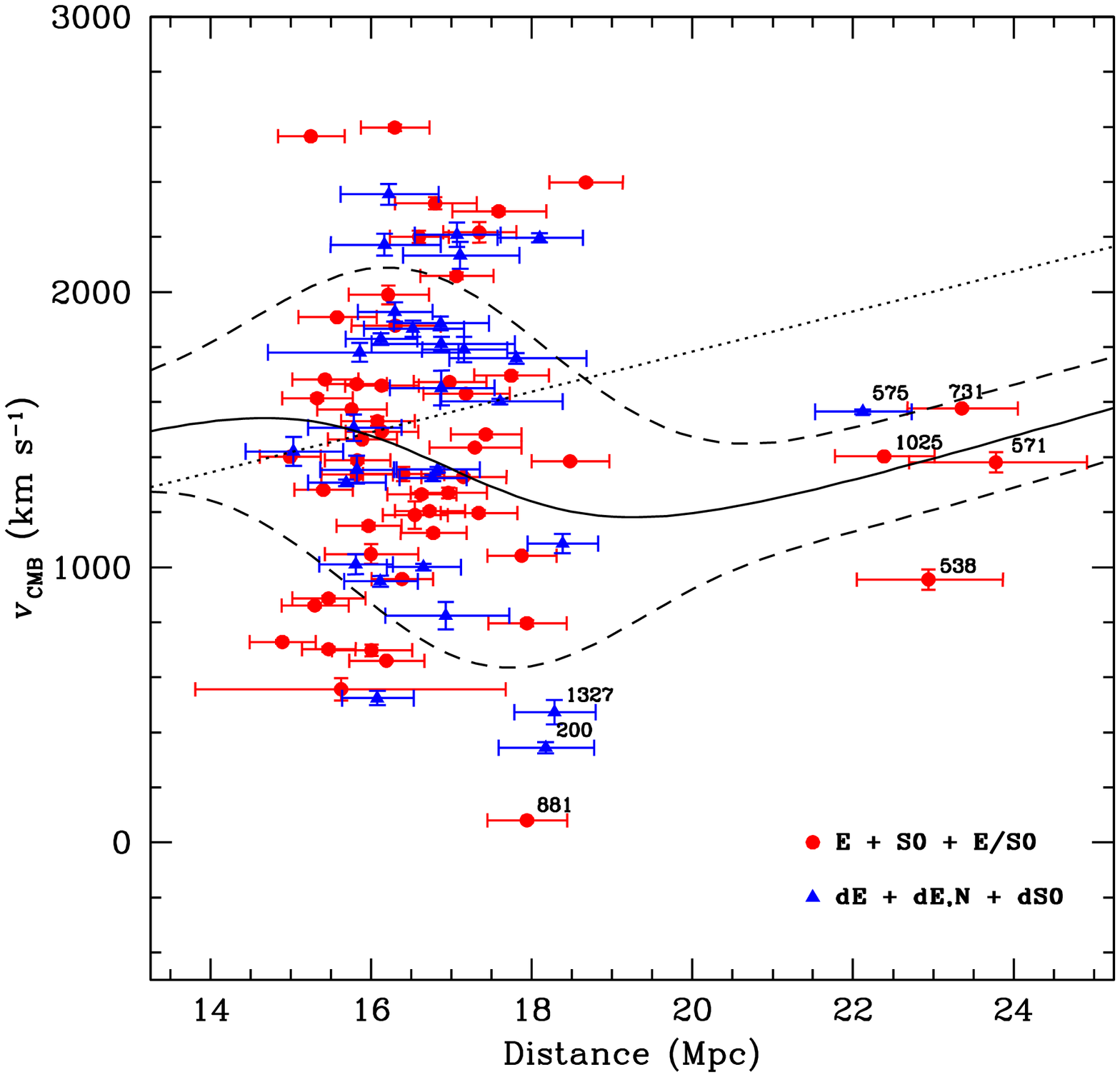}
\caption{Velocity-distance relation for galaxies from the ACS Virgo Cluster Survey.
Results for giants and dwarfs in the ACS Virgo Cluster Survey are
shown separately as the red circles and blue triangles, respectively.
The dotted lines shows the undisturbed Hubble Flow in the direction of Virgo for
an assumed Hubble Constant of $H_0 = 73$ \kms~Mpc$^{-1}$.
The predicted distance--velocity relation for a line of sight passing through the
cluster, based on the model of Tonry \etal (2000) for large scale flows in the
local universe, is shown by the solid (mean velocity) and dashed curves
($\pm 1\sigma$ limits). Note the presence of five galaxies associated with the
W$^{\prime}$ Cloud at $D \approx 23$~Mpc. A third grouping of galaxies 
associated with VCC881 (M86 = NGC4406) may also be present,
infalling from behind with a large negative radial velocity 
relative to the cluster mean.
\label{hubble1}}
\end{figure*}

Figure~\ref{hubble1} shows a Hubble diagram for the ACSVCS sample, after transforming
the heliocentric velocities into the CMB frame. Giants and dwarfs are shown separately
as red circles and blue triangles. The five members of the \wprime group are labeled
explicitly, as are three galaxies (VCC881, VCC200 and VCC1327) which may belong to the
third association mentioned above. The straight dotted line drawn through the main concentration 
of galaxies shows an unperturbed Hubble Flow for $H_0 = 73$~kms~Mpc$^{-1}$. The wavy
solid curve passing through the cluster is the large-scale flow model of Tonry \etal (2000)
for a line of sight passing through the Virgo Cluster. 
This model includes the contribution of the Virgo Cluster
(as well as the Great Attractor and a residual quadrupole) 
to the unperturbed Hubble flow.

\begin{figure}
\epsscale{1.2}
\plotone{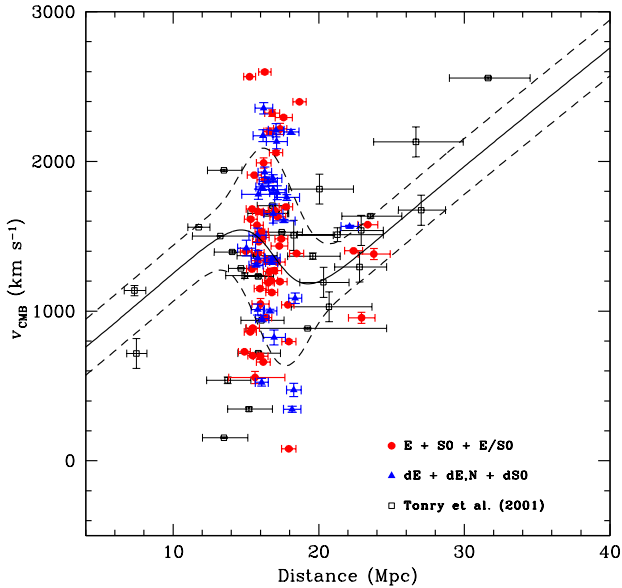}
\caption{Expanded view of the Hubble diagram shown in the previous figure. Over the 
distance range 4--40 Mpc, we plot giant and dwarf galaxies with measured SBF distances
from the ACS Virgo Cluster Survey (red circles and blue triangles). Early-type galaxies
from the SBF survey of Tonry \etal (2001) which are located with 20$^\circ$ of M87,
and which do not appear in the ACSVCS, are plotted as open squares. 
\label{hubble2}}
\end{figure}

The data and model are in good
accord, particularly for the \wprime Cloud which has the expected infall
velocity for its position. Moreover, the $rms$ scatter about the mean relation is well
described by the $\pm~1\sigma$ limits in the flow model (dashed curves).
Figure~\ref{hubble2} shows an expanded view of the region around the Virgo Cluster
(i.e., over the range 4--40~Mpc), in which the ACSVCS data have been augmented by 
galaxies from the Tonry \etal (2001) survey which lie within 20$^{\circ}$ of M87
and which were {\it not} included in the ACSVCS.\footnote{See Figure~21 of Tonry \etal
(2000) for an illustration of how the predicted Hubble flow relation depends on
direction.} There is good agreement with the flow model on this
larger scale as well.

\begin{figure}
\epsscale{1.2}
\plotone{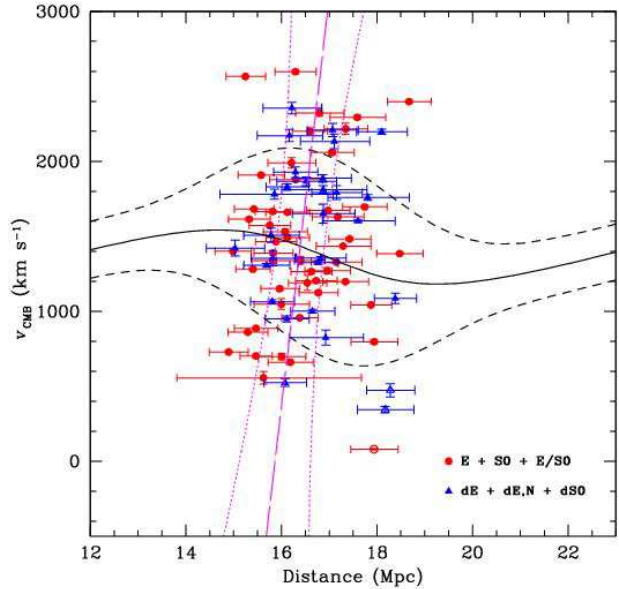}
\caption{Hubble diagram for the ACS Virgo Cluster Survey.
The long-dashed line shows a linear least squares fit to the measured velocities
and distances, excluding five members of the \wprime Cloud (not
shown) and the three galaxies (shown as open symbols) that may
belong to the VCC881 subcluster (i.e., 76 galaxies in total). The 2$\sigma$ confidence
bands on the fitted relation are shown by the dotted curves.
\label{hubble3}}
\end{figure}

Before proceeding, we comment on a curious feature of Figure~\ref{hubble2}. The
distribution in the distance-velocity plane seems to be slightly inclined, in the
sense that galaxies at high velocity tend to have larger distances 
while galaxies at low velocity tend to be on the near side of the cluster
(excepting M86 and its likely companions). Such a
correlation would be expected if the cluster has not yet virialized. Figure~\ref{hubble3}
shows a magnified view of the ACSVCS galaxies (with the \wprime Cloud galaxies
removed). A regression analysis of the remaining 79 galaxies shows that the inclination is
not statistically significant. However a marginally significant correlation ($\sim\,$2$\,\sigma$)
is found if the three galaxies that may belong to the M86 (VCC881) subcluster are omitted.
The resulting best-fit linear relation and its 2$\sigma$ confidence bands are shown as the 
long-dashed and dotted curves in Figure~\ref{hubble3}.  If we also omit the 2 highest
velocity galaxies, and thus fit the 74 main cluster galaxies in the velocity
range $500<v_{\rm CMB}<2500$ \kms, then the significance reaches 2.6$\,\sigma$, or 99\%.
This tentative correlation may be an echo of the 
cluster velocity distribution not yet having completely virialized. 
In the future, it will be interesting
to test this possibility using SBF measurements for an expanded sample of galaxies.

\subsection{The Principal Axis of the Virgo Cluster}
\label{pa}

Using SBF distances for 14 bright ellipticals from the catalog of Tonry \etal (2000, 2001),
West \& Blakeslee (2000) found that the Virgo Cluster has a ``principal axis" that extends
roughly along the line of sight toward Abell~1367, part of the Coma supercluster, at
$\alpha_{2000} \approx$ 11$^{\rm h}$44.5$^{\rm m}$,
$\delta_{2000} \approx 19.8^{\circ}$, $cz \approx 6600$ \kms and $d \approx 90$~Mpc
(e.g., Struble \& Rood 1999; Sakai et~al. 2000; Tully \& Pierce 2000).
The existence of a
preferred axis for Virgo's bright elliptical galaxies had previously been noted by Arp 
(1968) and Binggeli \etal (1987). In this section, we follow the analysis of 
West \& Blakeslee (2000) using an expanded sample of more precise SBF distances, with
the aim of determining the three-dimensional distribution of early-type Virgo galaxies,
as traced by the ACSVCS sample. We remind the reader at the outset that the ACSVCS
sample was chosen to give an unbiased view of the central ${\sim\,}5^{\circ}$ of the
Virgo Cluster, but it excludes galaxies associated with some of the more
interesting large-scale structures in the vicinity of Virgo 
(e.g., the Southern Extension, Ursa Major group, etc). 
The shape and orientation that we will find in our analysis
depend on the area coverage of our sample, and will be due
to the all the substructures included in it.

\begin{figure}
\epsscale{1.2}
\plotone{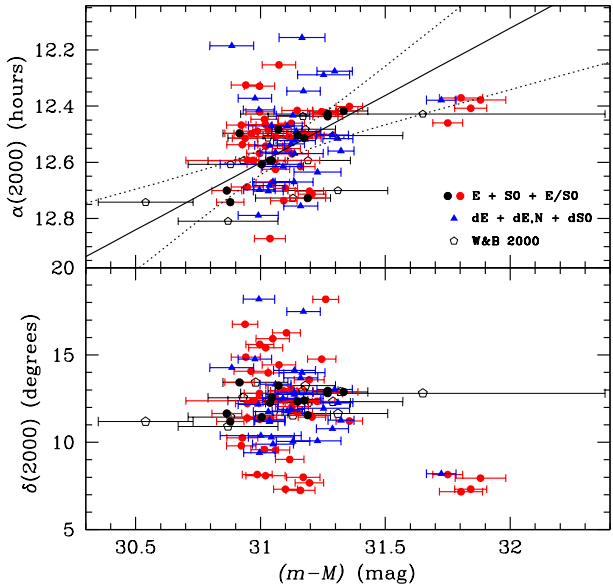}
\caption{{\it (Upper Panel}) Distance modulus plotted against right ascension for galaxies
from the ACS Virgo Cluster Survey. Dwarf and giant galaxies are shown by the blue triangles
and red circles, respectively. Open pentagons show the sample of galaxies (with SBF
measurements from Tonry \etal 2001) used by West \& Blakeslee (2000) to define Virgo's
principal axis; our measurements for these galaxies are shown as the black circles.
The straight line and dotted curves show the best-fit linear relation and 2$\sigma$
confidence bands based on our SBF distances for these galaxies.
{\it (Lower Panel}) Distance modulus plotted against declination for galaxies from the
ACS Virgo Cluster Survey. The symbols are the same as in the above panel.
\label{slice}}
\end{figure}

As a starting point, right ascensions, declinations and distances were used to define the
Cartesian coordinates, in Mpc, for each of the 84 galaxies with SBF distances. The
intrinsic standard deviations of the galaxy distribution (as defined in section 4.1) 
about these respective axes is then found to be
$\sigma_{\rm o}({\alpha}) = 0.55 \pm 0.05$~Mpc, $\sigma_{\rm o}({\delta}) = 0.67 \pm 0.07$~Mpc, and
$\sigma_{\rm o}(d) = 1.62 \pm 0.41$~Mpc. Discarding the five galaxies at $d \approx 23$~Mpc
which are almost certainly members of the W$^{\prime}$ cloud, one finds
$\sigma_{\rm o}({\alpha}) = 0.55 \pm 0.05$~Mpc, $\sigma_{\rm o}({\delta}) = 0.62 \pm 0.06$~Mpc, and
$\sigma_{\rm o}(d) = 0.62 \pm 0.06$~Mpc. 
Figure~\ref{slice}
plots distance modulus against right ascension and declination for the ACSVCS
galaxies (red circles and blue triangles). The open pentagons show the SBF measurements
from Tonry \etal (2001) which were used by West \& Blakeslee (2002) to show that the distance
moduli of the brightest ellipticals vary with right ascension. This trend is not as obvious
with our new sample, although a least-squares fit (shown as the straight line in the upper
panel) to our new measurements for the same sample of (bright) galaxies reveals a correlation
consistent with the West \& Blakeslee (2000) data. The dotted curves show the 2$\sigma$
confidence bands on the fitted relation.
It is not surprising that the correlation between distance modulus and right ascension is more diluted in our sample, since it includes galaxies over
a much larger area than in West \& Blakeslee (2000).

To better quantify the shape of the early-type galaxies in Virgo, we have calculated
the inertia tensor of the galaxy distribution with respect to the cluster's
center of mass. We have adopted as the fixed centroid of our galaxy distribution
the point corresponding to average right ascension, declination, and distance (calculated by excluding the galaxies with  $(m-M) > 31.5 $mag).
The contributions to the scatter from observational error and cosmic variance
can then be easily subtracted from the diagonal terms of the tensor to derive 
the intrinsic spatial distribution.
The uncertainties in the distances are uncorrelated with the three positional variables.
The principal axes of the galaxy distribution are then given by the eigenvectors of the
inertia tensor; these axes identify the major, minor and intermediate axes of the 
ellipsoid that best fits the galaxy distribution. The elongation along these
axes are given by the eigenvalues of the tensor.

In Table~\ref{shape}, shape parameters for the galaxy distribution are shown as the major,
intermediate and minor axis ratios --- {\it a,b,} and {\it c}, respectively --- after
normalizing to the major axis. These values and their uncertainties were obtained
by bootstrapping on 1000 simulations of each of our samples.
As an illustration, with this parameterization, a straight
line would have ($a,b,c$) = (1,~0,~0) and a sphere would have (1,~1,~1).
Excluding the five probable members of the W$^{\prime}$ Cloud, which have $(m-M) > 31.5$~mag,
gives axis ratios of (1.0,~0.7,~0.5). These values are representative of those found for
various subsets of the 79 galaxies with $(m-M) < 31.5$~mag, after dividing the sample on
the basis of color, luminosity and/or morphological type (i.e., giants versus dwarfs). 
If one chooses to weight the sample by luminosity, the ratios typically change by 
$\sim$~30\% percent, approaching the distribution of the brightest ($M_B < 12$~mag) 
galaxies, which has a more elongated distribution with respect to the total distribution.

\begin{figure}
\epsscale{1.2}
\plotone{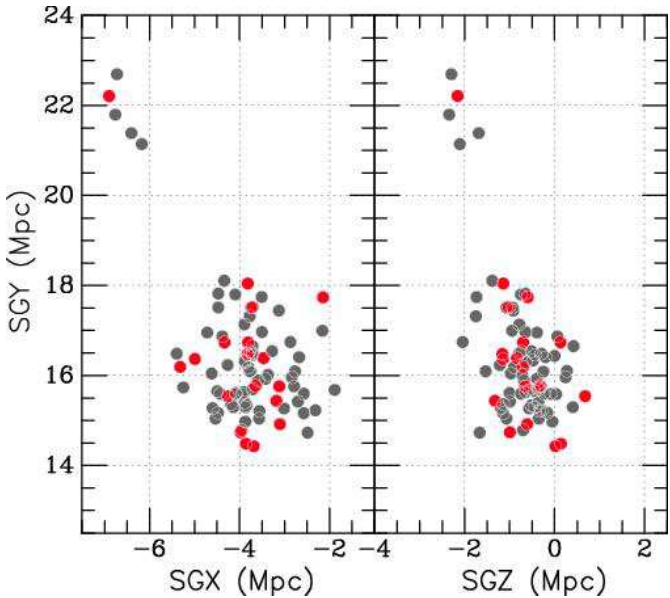}
\caption{The ACSVCS galaxy spatial
distribution plotted in Supergalactic Cartesian coordinates, with the SGY
axis being approximately along the line of sight.
The left panel shows the view from ``above'' the supergalactic plane,
with the SGY axis being along the line of sight.  The right panels give
the view within the plane. 
The brightest ellipticals ($B_T < 12$~mag),  are plotted in red.
\label{sgc}}
\end{figure}

Figure~\ref{sgc} shows the sample distribution in supergalactic coordinates,
with red circles indicating the brightest galaxies in the sample ($M_B < 12$~mag).
Supergalactic coordinates have their equator aligned with the supergalactic plane, 
defined by de Vaucouleurs (1991) as the plane in which most of the structures
 belonging to the local supercluster (centered on the Virgo cluster) lie. 
In particular, the origin of the spherical supergalactic coordinate system
($SGB=0^{\circ}, SGL=0^{\circ}$)
 lies at galactic coordinates $l=137.37^{\circ}, b=0^{\circ} $;
that is $\alpha(2000) \sim 42^{\circ}$ and
$\delta(2000) \sim +59.5^{\circ}$, with the north pole at galactic coordinates 
$l=47.37^{\circ}, b=+6.32^{\circ} $.
We projected the spherical supergalactic coordinates to a Cartesian plane.
The galaxy distribution is significantly flattened in the SGZ direction.

In summary, the distribution of our sample of early-type Virgo galaxies appears 
triaxial, although with only a mild elongation along the major axis. This conclusion
is largely independent of galaxy color and morphological classification, with the dwarfs and giants
having similar triaxial distributions. 
As shown in Table~\ref{shape}, the principal axis of the early-type galaxy distribution is found to 
be inclined at an angle of $\sim$ 20-70$^{\circ}$ to the line of sight to the cluster (i.e.,
elongated roughly along the direction to the cluster), with the precise inclination depending
on the sample definition. In particular, if the sample is not 
restricted to exclude the five members of the background W$^{\prime}$ Cloud, then a much more
elongated distribution is found: i.e.,  (1,~0.4,~0.3), inclined at an angle of $\sim$ 10$^{\circ}$.
The background W Cloud, at roughly twice the distance of Virgo, appears to connect to the
cluster via a filament (that includes
the \wprime Cloud) wrapping around a background ``void" and connecting  with the
Virgo Southern Extension (R.~Tully, private communication).
A larger elongation is also observed for the brightest galaxies ($M_B < 12$~mag), 
at an an angle of 
$\sim$ 30$^{\circ} \pm 20^{\circ}$ to the line of sight.

\section{Conclusions}

\begin{figure*}
\epsscale{0.8}
\plotone{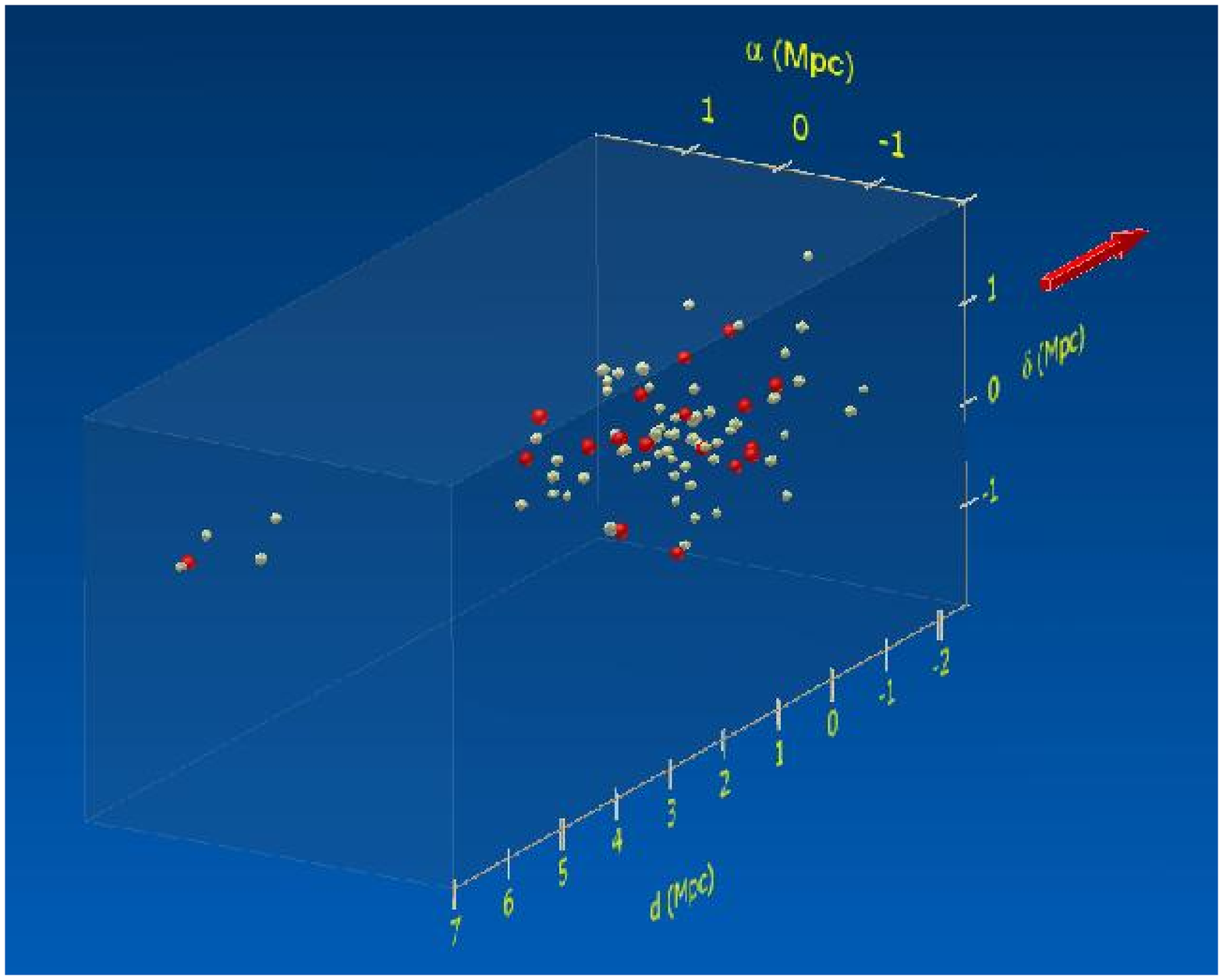}
\caption{Three-dimensional distribution of our program galaxies within the Virgo Cluster. 
The cluster is embedded in a rectangular parallelepiped of dimensions $4\times4\times9.5$~Mpc.
The red spheres show galaxies with $B_T \le 12$~mag. The direction of the Milky Way is
indicated by the arrow.
\label{virgo3d}}
\end{figure*}

SBF distances from the ACS Virgo Cluster Survey offer the best opportunity to date to map out
the three-dimensional distribution of early-type galaxies within the Virgo Cluster. The final
sample of 84 galaxies (50 giants and 34 dwarfs) with SBF distances from this survey nearly triples the number 
of Virgo Cluster galaxies with available SBF distances. At the same time, the new $z_{850}$-band
SBF magnitudes and distances have a typical  (internal) precision of $\approx$ 0.07~mag and
$\approx$ 0.5~Mpc --- roughly a  factor of three improvement over previous measurements.
An illustration of the three-dimensional distribution of our program galaxies within the Virgo
Cluster is presented in Figure~\ref{virgo3d}.

Five galaxies in the survey (VCC538, VCC571, VCC575, VCC731, VCC1025) lie well behind the
Virgo Cluster, at a mean distance of $d \approx 22.9\pm0.3$~Mpc.  These galaxies are almost
certainly members of the \wprime Cloud. Binggeli \etal (1993) have previously described this
structure as a filament, viewed nearly end-on, that connects the Virgo B subcluster with the
background W Cloud (at roughly twice the distance of the Virgo Cluster). Our finding that
these five galaxies occupy such an apparently narrow range in distance (to within the errors,
there is no evidence for any spread in distance at all) suggests that  the \wprime Cloud may
be a more localized structure than previously believed.

Excluding these background \wprime\ galaxies,
we find the remaining sample of 79 galaxies to occupy a narrow range in 
distance, with a mean of $d = 16.5\pm0.1$~(random) $\pm1.1$~Mpc (systematic).
After accounting for measurement errors and cosmic scatter, we find the 1$\sigma$ dispersion
in distance to be $\sigma(d) = 0.6\pm0.1$~Mpc. Our estimate for the back-to-front depth of
the cluster is then $\approx 2.4\pm0.4$~Mpc (i.e., $\pm2\sigma$ of the distance distribution: a
range that should encompass 95\% of the cluster's early-type galaxies). This finding is clearly
at odds with some claims of much more elongated distributions found using less precise
distance indicators (e.g., Young \& Currie 1995). At the same time, though, our carefully chosen
sample turned out to include five members of the \wprime Cloud, despite the fact that all were
considered certain Virgo members in the VCC (Binggeli \etal 1985). 
Although ground-based SBF measurements (Tonry \etal\ 1990; 
Tonry \etal\ 2001) had already placed NGC\,4365 (VCC731) and NGC\,4434 (VCC1025)
in the background towards the direction of the W~Cloud, 
the existence of such a compact, distinct physical grouping displaced by 
$\sim$~6~Mpc from the main body of the cluster would almost certainly
have gone unrecognized in studies using distances of precision $\sim$ 3~Mpc.

We therefore suspect
that contamination by galaxies lying in the immediate cluster background may at least partly
account for the elongated distributions reported by some previous researchers. 
We also note that, due to our sample definition, these background
galaxies would not have been included in the sample if the distance offset
were perpendicular to the line of sight (as is the case for the Virgo
Southern Extension), rather than along it.  
Parsing the sample of early-type galaxies by morphological type, luminosity and color reveals no 
large differences in distribution, although there is some evidence that the giants may have a slightly
more elongated distribution 
than the dwarfs ($0.7\pm0.1$~Mpc versus $0.5\pm0.1$~Mpc). We emphasize that these measurements 
are based on {\it early-type} galaxies, which are known to be
more concentrated toward the cluster center
than the late-type systems. Our SBF measurements show that the well known
A and B  subclusters (defined by M87 and M49, respectively)  lie at a nearly identical distance
of $d \approx 16.5$~Mpc.  A small number of the galaxies in our survey may belong
to a third subcluster, about 1~Mpc more distant,
associated with M86 (see also Schindler \etal 1999; Jerjen \etal 2004). 

A tensor of inertia analysis is used to examine the three-dimensional distribution of early-type
galaxies in the Virgo Cluster. We find the main body of the
cluster to have a mildly triaxial shape, with axial ratios of (1,~0.7,~0.5).
If the five \wprime\ Cloud galaxies are included in the solution, then the axial ratios become
(1,~0.4,~0.3), because these galaxies don't lie in the center of the cluster
as most of the galaxies of the sample.
The principal axis of this distribution is inclined by $\sim$20-40$^{\circ}$
from the line of sight. 

While our conclusions are robust to the precise choice of SBF calibration (see Appendix~A), in
the future we
plan to refine our adopted calibration using SBF measurements for a  sample of $\sim$ 40 
galaxies in the Fornax Cluster (e.g., Jord\'an \etal 2005b about our Fornax Cluster Survey (FCS); Dunn \& Jerjen 2006 about the use of SBF in measuring distances in the Fornax cluster). Because it is much more
compact than Virgo (Tonry et al. 1991), the SBF magnitudes measured in Fornax galaxies will allow a more 
direct and straightforward calibration of the ${\overline M_{850}}-(g_{475}-z_{850})_0$ relation.  
A forthcoming paper in this series will use our SBF distances to re-assess the accuracy of 
several distance indicators commonly used to derive distances for early-type galaxies.

\acknowledgments
Support for program GO-9401 was provided through a grant from the Space
Telescope Science Institute, which is operated by the Association of 
Universities for Research in Astronomy, Inc., under NASA contract NAS5-26555. 
ACS was developed under NASA contract NAS 5-32865.
S.M.  acknowledges additional support from NASA grant 
NAG5-7697 to the ACS Team, and from NASA JPL/Spitzer grant RSA 1264894.
P.C. acknowledges support provided by NASA LTSA grant NAG5-11714.
M.J.W. acknowledges support through NSF grant AST-0205960.
D.M. acknowledges support provided by NSF grants AST-0071099, AST-0206031, 
AST-0420920 and AST-0437519, by NASA grant NNG04GJ48G, and by grant 
HST-AR-09519.01-A from STScI. 
This research has made use of the NASA/IPAC Extragalactic Database (NED)
which is operated by the Jet Propulsion Laboratory, California Institute
of Technology, under contract with the National Aeronautics and Space Administration.
We thank R.~B. Tully and B. Binggeli for their insightful comments, and the anonymous
referee for a careful reading of the manuscript and many helpful comments that
improved the clarity of the presentation.

\appendix{

\section{Choice of Calibration}

To understand the uncertainties inherent to the choice of SBF calibration, we compare the
results obtained using the broken-linear relation of Equation~\ref{eq:broken} with
two other possible calibrations: a single-slope linear fit and a fourth-order polynomial fit.

The calibration by Equation~\ref{eq:broken} was chosen to minimize the $\chi^2$ of a fit
of SBF measurements as a function of $(g_{475}-z_{850})_0$ for 84 galaxies, in two color
regimes divided at $(g_{475}-z_{850})_0 = 1.3$~mag. Using the same sample and the least-squares
fitting technique, the best-fit linear calibration  is
\begin{equation}
\begin{array}{rrrrr}
\overline M_{850} & = & -2.00 \pm 0.04 + (1.3 \pm 0.1)[ (g_{475}-z_{850})_0-1.3 ] & & 1.0 \le (g_{475}-z_{850})_0 \le 1.6 \\
\end{array}
\label{eq:linear}
\end{equation}
while a fourth-order polynomial fit gives
\begin{equation}
\begin{array}{rrrrr}
\overline M_{850} & = & -2.04 \pm 0.05 + (1.12 \pm 0.36)x + (1.97 \pm 2.00)x^2 \\
&   & + (8.38 \pm 8.28)x^3 + (11.33 \pm 23.46)x^4  & 1.0 \le (g_{475}-z_{850})_0 \le 1.6 \\
\end{array}
\label{eq:poly}
\end{equation}
with $x= [(g_{475}-z_{850})_0-1.3]$. In Equation~\ref{eq:poly}, we approximate the errors on
the polynomial coefficients as the square root of the diagonal terms of the fit covariance
matrix.  Distance moduli derived from the three different calibrations are presented in
Table~\ref{tabledm}, and galaxy-by-galaxy differences in distance modulus obtained using
Equations~\ref{eq:linear} or Equation~\ref{eq:poly}, rather than Equations~\ref{eq:broken},
are shown in the upper panel of Figure~\ref{calib2}. For reference, the lower panel of
this figure shows the adopted distance moduli plotted against galaxy color. There is no 
trend with either color or morphological type, supporting the validity of the broken 
linear calibration adopted in \S\ref{sec:cal}.

In performing these fits, we add two additional sources of uncertainty to the errors on the
SBF magnitudes. The first is an ``uncertainty" of 0.11~mag, introduced to account for the actual
distance dispersion caused by the line-of-sight depth of the cluster. This value was adopted
based on the $rms$ scatter of the angular distances measured within $\sim3^\circ$ in position
on the sky of the 100 ACSVCS program galaxies. The second uncertainty is the expected 
``cosmic scatter'' of 0.05~mag in fluctuation magnitude (see Tonry \etal 1997), representing 
the intrinsic dispersion in $\overline M_{850}$ for galaxies at a fixed color.

The three different calibrations are shown in the lower panel of Fig.~\ref{calib1}, and their
normalized $\chi^2$ and scatters are given in Table~\ref{cali} for the 79 galaxies having
distance moduli of $(m-M) < 31.5$~mag. Note that the normalized $\chi^2$ are larger than
expected when we do not include the uncertainties due to depth effects in the Virgo Cluster
and the intrinsic dispersion in $\overline M_{850}$. When these sources of scatter are taken
into account, $\chi^2$ has values around unity. Although the $\chi^2$ for the polynomial fit
is slightly lower when the full galaxy sample is considered, it increases for the subset of
galaxies with $(g_{475}-z_{850})_0 < 1.5$~mag. This is because the polynomial calibration 
gives a better match to those few galaxies with $(g_{475}-z_{850})_0 > 1.5$~mag.
\begin{figure}
\epsscale{0.7}
\plotone{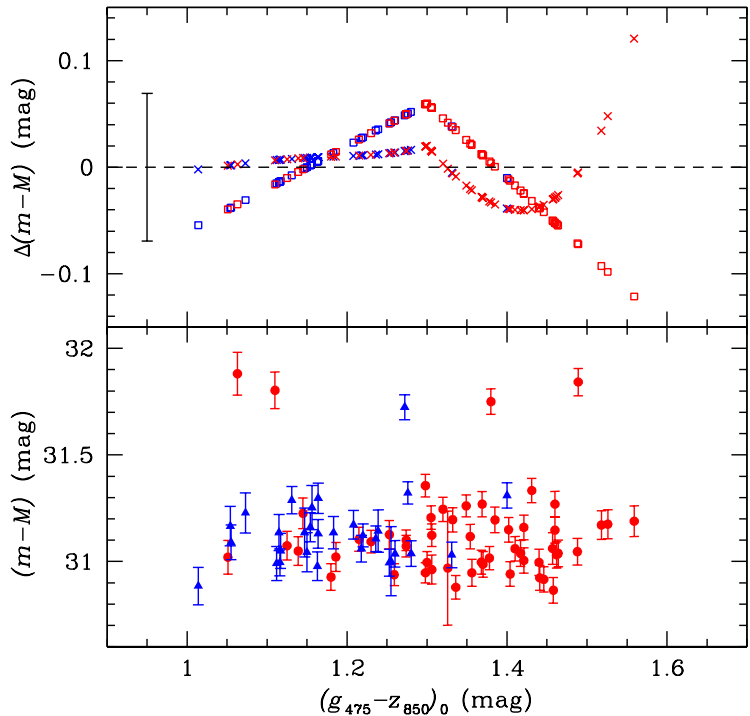}
\caption{{\it (Upper Panel)} Differences between distance moduli derived with
the broken linear calibration (Equation~\ref{eq:broken}) and the single linear
(squares; Equation~\ref{eq:linear}) and polynomial calibrations (crosses;
Equation~\ref{eq:poly}). Red and blue symbols show galaxies classified as giants
and dwarfs, respectively, by Binggeli \etal (1985). The errorbar indicates the
mean random error, $\sigma(m-M) = 0.069$~mag, on the measured distance moduli.
{\it (Lower Panel)} Distance moduli as a function of galaxy color. 
Red circles and blue triangles are used for giants and dwarfs, respectively.
The distance moduli show no correlation with color, supporting the validity
of the adopted SBF calibration.
\label{calib2}}
\end{figure}
In fact, the three brightest and reddest galaxies in our sample --- all with
$(g_{475}-z_{850})_0 > 1.5$~mag --- have SBF magnitudes slightly larger (but
always within 1~$\sigma$) than expected if they lie at the precise center of the
cluster. As a consequence, Equation~\ref{eq:broken} yields distance moduli which
are $\approx$~0.08~mag larger than the mean Virgo distance for these three
galaxies (M87, M49 and M60).  While this modest offset may represent a real
dislocation of these three galaxies from the cluster center, it may also be an 
artifact of the choice of SBF calibration, i.e., the steepening of the SBF--color
relation for the very reddest galaxies may be real.  The composite stellar
population models presented by Blakeslee \etal\ (2001) showed such a steepening
of the $I$-band SBF relation for very metal-rich populations (see also Fig.~7 of
Paper~V, showing Bruzual \& Charlot 2003 SSP models for the F850LP
bandpass).  In any case, the average uncertainty in distance modulus is
0.07~mag, so that distance moduli derived using the different calibrations are
always consistent with the 1$\sigma$ errors, with the exception of VCC1978 (M60)
--- the reddest galaxy in the sample. Nevertheless, we advise future users of
the SBF distance catalog to exercise caution by comparing all three distance
sets to test for robustness against the choice of calibration.

}

\newpage

\clearpage

\LongTables
\begin{deluxetable}{crcccccccl}
\tabletypesize{\scriptsize}
\tablecaption{SBF Distances for ACS Virgo Cluster Survey Galaxies.\label{catalogue}}
\tablewidth{0pt}
\tablehead{
\colhead{ID} & 
\colhead{VCC} &
\colhead{($g_{475}-z_{850})_0$} &
\colhead{$\overline m_{850}$} &
\colhead{(m-M)} &
\colhead{$B_T^b$} &
\colhead{$\langle v_r\rangle$} &
\colhead{Type} &
\colhead{Other}& 
\colhead{Note} \\
\colhead{No.} & 
\colhead{No.} & 
\colhead{(mag)} &
\colhead{(mag)} &
\colhead{(mag)} &
\colhead{(mag)} &
\colhead{(km~s$^{-1}$)} &
\colhead{} &
\colhead{} &
\colhead{}
}
\startdata
  1 & 1226 &1.52  $\pm$  0.01&      29.55  $\pm$   0.01&      31.17  $\pm$     0.07 & 9.31 &  997$\pm$7 & E2/S0$_1$(2)  & M49,~N4472 & \\
  2 & 1316 &1.53  $\pm$  0.01&      29.57  $\pm$   0.01&      31.18  $\pm$     0.07 & 9.58 & 1307$\pm$7 & E0            & M87,~N4486 & \\
  3 & 1978 &1.56 $\pm$ 0.01&29.65 $\pm$ 0.01&  31.19  $\pm$     0.07&  9.81 & 1117$\pm$6 & S0$_1$(2)     & M60,~N4649  & \\
  4 &  881 &1.46 $\pm$ 0.01 &29.53  $\pm$ 0.04&       31.13  $\pm$     0.07&10.06 & -244$\pm$5 & S0$_1$(3)/E3  & M86,~N4406 &  \\
  5 &  798 &1.35  $\pm$ 0.01 &29.30 $\pm$ 0.01&      31.26  $\pm$     0.05 & 10.09 &  729$\pm$2 & S0$_1$(3)~pec & M85,~N4382 &  \\
  6 &  763 &1.43  $\pm$ 0.01 &29.54 $\pm$ 0.02& 31.33 $\pm$ 0.05& 10.26  & 1060$\pm$6 &E1            & M84,~N4374 &  \\
  7 &  731 &1.49 $\pm$ 0.01&30.16 $\pm$ 0.02&   31.84 $\pm$ 0.06& 10.51  &1243$\pm$6 &  E3            & N4365 & \\
  8 & 1535 & \nodata             & \nodata & \nodata & 10.61 &   448$\pm$8 & S0$_3$(6)     & N4526 & 1 \\
  9 & 1903 &1.46 $\pm$ 0.01&  29.12  $\pm$  0.01&      30.87  $\pm$     0.06   & 10.76 &  410$\pm$6 & E4            & M59,~N4621 &  \\
 10 & 1632 &1.49 $\pm$ 0.01& 29.19  $\pm$ 0.02&      31.00  $\pm$     0.06   & 10.78 &  340$\pm$4 & S0$_1$(0)     & M89,~N4552 &  \\
 11 & 1231 &1.45  $\pm$     0.01&      29.15  $\pm$   0.02&      30.92  $\pm$     0.06  & 11.10 & 2244$\pm$2 & E5            & N4473  & \\
 12 & 2095 & \nodata &\nodata& \nodata& 11.18 &  984$\pm$11& S0$_1$(9)     & N4762  & 2\\
 13 & 1154 & 1.46  $\pm$     0.01&      29.29  $\pm$   0.02&      31.03  $\pm$     0.06& 11.37 & 1210$\pm$16& S0$_3$(2)     & N4459 & \\
 14 & 1062 & 1.44  $\pm$     0.01&      29.15  $\pm$    0.02&      30.92  $\pm$     0.06  & 11.40 &  532$\pm$8 & SB0$_1$(6)    & N4442 &  \\
 15 & 2092 &  1.46  $\pm$     0.01&      29.31  $\pm$   0.02&      31.04  $\pm$     0.06 & 11.51 & 1347$\pm$9& SB0$_1$(5)    & N4754 &  \\
 16 &  369 & 1.44  $\pm$     0.01&      29.22  $\pm$    0.02&      31.00  $\pm$     0.06    & 11.80 & 1009$\pm$13& SB0$_1$       & N4267 & \\
 17 &  759 &  1.46  $\pm$     0.01&      29.41  $\pm$  0.02&      31.15  $\pm$     0.06 & 11.80 &  943$\pm$19& SB0$_2$(r)(3) & N4371 &  \\
 18 & 1692 & 1.42  $\pm$     0.01&      29.34  $\pm$   0.02&      31.16  $\pm$     0.06    & 11.82 & 1730$\pm$13& S0$_1$(7)/E7  & N4570 & \\
 19 & 1030 & 1.31  $\pm$     0.01&      29.08  $\pm$    0.02&      31.12  $\pm$     0.05 & 11.84 &  801$\pm$10& SB0$_1$(6)    & N4435 &  \\
 20 & 2000 &1.34  $\pm$     0.01&      28.89  $\pm$   0.02&      30.88  $\pm$     0.05& 11.94 & 1083$\pm$4 & E3/S0$_1$(3)  & N4660 & \\
 21 &  685 &\nodata&\nodata &\nodata  &11.99 & 1200$\pm$15& S0$_1$(8)     & N4350 & 1,2\\
 22 & 1664 &1.42  $\pm$     0.01&      29.19  $\pm$   0.02&      31.00  $\pm$     0.06    & 12.02 & 1142$\pm$2 & E6            & N4564 & \\
 23 &  654 &\nodata&\nodata&\nodata       & 12.03 &  950$\pm$9 & RSB0$_2$(5)   & N4340 & 2 \\
 24 &  944 & 1.38  $\pm$     0.01&      29.11  $\pm$   0.02&      31.02  $\pm$     0.06   & 12.08 &  820$\pm$12& S0$_1$(7)     & N4417 &  \\
 25 & 1938 &1.30  $\pm$     0.01&      29.16  $\pm$   0.02&      31.21 $\pm$     0.05  & 12.11 & 1164$\pm$10& S0$_1$(7)     & N4638 &  \\
 26 & 1279 &1.40  $\pm$     0.01&      29.29  $\pm$   0.02&      31.15  $\pm$     0.06 & 12.15 & 1349$\pm$3 & E2            & N4478 & \\
 27 & 1720 &1.41  $\pm$     0.01& 29.22 $\pm$ 0.02  & 31.06 $\pm$ 0.06 & 12.29 & 2273$\pm$12& S0$_{1/2}$(4) & N4578  & 5 \\
 28 &  355& 1.40  $\pm$     0.01&      29.09  $\pm$   0.02&      30.94  $\pm$     0.06   & 12.41 & 1359$\pm$4 & SB0$_{2/3}$   & N4262  & \\
 29 & 1619 &      1.30  $\pm$     0.01&      28.89  $\pm$   0.02&      30.95 $\pm$     0.05  & 12.50 &  381$\pm$9 & E7/S0$_1$(7)  & N4550 &   \\
 30 & 1883 & 1.28 $\pm$     0.01&      29.02  $\pm$   0.02&      31.10  $\pm$     0.05 & 12.57 & 1875$\pm$22& RSB0$_{1/2}$  & N4612 &  \\
 31 & 1242 & 1.31  $\pm$     0.01&      28.92  $\pm$   0.05&      30.96  $\pm$     0.07 & 12.60 & 1588$\pm$7& S0$_1$(8)     & N4474  & \\
 32 &  784 & 1.37  $\pm$     0.01&      29.07  $\pm$ 0.02&      31.00  $\pm$     0.06   & 12.67 & 1069$\pm$10& S0$_1$(2)     & N4379 &  \\
 33 & 1537 &  1.30  $\pm$     0.01&      28.94 $\pm$    0.03&      31.00  $\pm$     0.05  & 12.70 & 1342$\pm$8& SB0$_2$(5)    & N4528 &  \\
 34 &  778 & 1.32  $\pm$     0.01&      29.22  $\pm$   0.03&      31.25  $\pm$     0.06 & 12.72 & 1375$\pm$11& S0$_1$(3)     & N4377 & \\
 35 & 1321 &    1.26  $\pm$     0.01&      28.84  $\pm$     0.03&      30.94  $\pm$     0.05   & 12.84 &  967$\pm$6 & S0$_1$(1)     & N4489 &  \\
 36 &  828 & 1.37  $\pm$     0.01&      29.35  $\pm$   0.03&      31.27  $\pm$     0.06 & 12.84 &  472$\pm$11& E5            & N4387  & \\
 37 & 1250 &     1.14  $\pm$     0.01&      29.03  $\pm$  0.05&      31.23  $\pm$     0.07 & 12.91 & 1970$\pm$11& S0$_3$(5)     & N4476 &   \\
 38 & 1630 & 1.42  $\pm$     0.01&      29.21  $\pm$   0.03&      31.04  $\pm$     0.06   & 12.91 & 1172$\pm$6 & E2            & N4551 &  \\
 39 & 1146 &1.27  $\pm$     0.01&      28.99  $\pm$   0.03&      31.07  $\pm$     0.05 & 12.93 &  635$\pm$6 & E1            & N4458 &  \\
 40 & 1025 &1.38 $\pm$     0.02 & 29.85  $\pm$ 0.03 & 31.75  $\pm$ 0.06 & 13.06 & 1071$\pm$6 & E0/S0$_1$(0)  & N4434  & 5 \\
 41 & 1303 & 1.35  $\pm$     0.01&      29.16  $\pm$  0.03&      31.12  $\pm$     0.06 & 13.10 &  875$\pm$10& SB0$_1$(5)    & N4483 &  \\
 42 & 1913 & 1.33  $\pm$     0.01&      29.20  $\pm$  0.03&      31.20  $\pm$     0.06 & 13.22 & 1892$\pm$37& E7            & N4623  & \\
 43 & 1327 &1.40 $\pm$     0.01 & 29.45 $\pm$  0.03&  31.31  $\pm$     0.06   & 13.26 &  150$\pm$45& E2            & N4486A& 3 \\
 44 & 1125 &\nodata&\nodata &\nodata      & 13.30 &  165$\pm$6& S0$_1$(9)     & N4452 & 2 \\
 45 & 1475 & 1.21  $\pm$     0.01&      28.97  $\pm$   0.03&      31.10  $\pm$     0.06 & 13.36 &  951$\pm$11& E2            & N4515 &  \\
 46 & 1178 &  1.37  $\pm$     0.01&      29.07  $\pm$   0.03&      31.00  $\pm$     0.06   & 13.37 & 1243$\pm$2 & E3            & N4464 &  \\
 47 & 1283 & 1.38  $\pm$     0.01&      29.30  $\pm$   0.03&      31.20  $\pm$     0.06 & 13.45 &  876$\pm$10& SB0$_2$(2)    & N4479 &  \\
 48 & 1261 &1.13  $\pm$     0.01&      29.08  $\pm$     0.04&      31.29  $\pm$     0.06  & 13.56 & 1871$\pm$16& d:E5,N        & N4482 &   \\
 49 &  698 & 1.30 $\pm$     0.01&      29.29  $\pm$   0.03&      31.36  $\pm$     0.05  & 13.60 & 2070$\pm$7& S0$_1$(8)     & N4352  & \\
 50 & 1422 &1.18 $\pm$     0.01&      28.76  $\pm$    0.04&      30.93  $\pm$     0.06 & 13.64 & 1288$\pm$10& E1,N:         & I3468 &  \\
 51 & 2048 &\nodata&\nodata &\nodata      & 13.81 & 1084$\pm$12& d:S0(9)       & I3773 &  2 \\
 52 & 1871 & 1.36  $\pm$     0.01&      29.00  $\pm$  0.04&      30.95 $\pm$     0.06  & 13.86 &  567$\pm$10& E3            & I3653  & \\
 53 &    9 & 1.05  $\pm$     0.01&      28.89  $\pm$   0.07&      31.17 $\pm$     0.09 & 13.93 & 1804$\pm$49& dE1,N         & I3019  & \\
 54 &  575 &1.27$\pm$ 0.01 &29.64 $\pm$ 0.04 & 31.72 $\pm$ 0.06  & 14.14 & 1231$\pm$9 & E4            & N4318 &  \\
 55 & 1910 &  1.33  $\pm$     0.01&      29.03  $\pm$   0.03&      31.03  $\pm$     0.06  & 14.17 &  206$\pm$26& dE1,N         & I809  &  \\
 56 & 1049 & 1.05  $\pm$     0.01&      28.74  $\pm$  0.05&      31.02  $\pm$     0.08  & 14.20 &  716$\pm$36& S0(4)         & U7580 &  \\
 57 &  856 & 1.16  $\pm$     0.01&      28.95  $\pm$ 0.04&      31.13  $\pm$     0.07  & 14.25 & 1025$\pm$10& dE1,N         & I3328 & \\
 58 &  140 &1.13  $\pm$     0.01&      28.86  $\pm$  0.04&      31.07  $\pm$     0.07   & 14.30 & 1013$\pm$26& S0$_{1/2}$(4) & I3065 & \\
 59 & 1355 &  1.12  $\pm$     0.01&      28.91  $\pm$   0.06&      31.14  $\pm$     0.08 & 14.31 & 1332$\pm$63 & dE2,N         & I3442  &  \\
 60 & 1087 &1.24  $\pm$     0.01&      28.99  $\pm$   0.04&      31.11  $\pm$     0.06   & 14.31 &  675$\pm$12& dE3,N         & I3381  & \\
 61 & 1297 & 1.46  $\pm$   0.01&      29.31  $\pm$   0.04&      31.06  $\pm$     0.07 & 14.33 & 1555$\pm$4 & E1            & N4486B & \\
 62 & 1861 &  1.26  $\pm$     0.01&      28.94  $\pm$  0.05&      31.04  $\pm$     0.06 & 14.37 &  629$\pm$20& dE0,N         & I3652  & \\
 63 &  543 & 1.16  $\pm$     0.01&      28.79  $\pm$   0.05&      30.98  $\pm$     0.07  & 14.39 &  985$\pm$12& dE5           & U7436  & \\
 64 & 1431 &1.28  $\pm$     0.01&      28.96  $\pm$   0.04&      31.04  $\pm$     0.06   & 14.51 & 1505$\pm$21& dE0,N         & I3470  & \\
 65 & 1528 & 1.22  $\pm$     0.01&      28.93  $\pm$   0.04&      31.06 $\pm$     0.06 & 14.51 & 1608$\pm$35& d:E1          & I3501   & \\
 66 & 1695 &   1.05  $\pm$     0.01&      28.81  $\pm$  0.05&      31.09  $\pm$     0.08  & 14.53 & 1547$\pm$29& dS0:          & I3586   & \\
 67 & 1833 &  1.14  $\pm$     0.01&      28.84  $\pm$   0.04&      31.05  $\pm$     0.07 & 14.54 & 1679$\pm$34& S0$_1$(6)     &         & \\
 68 &  437 & 1.21  $\pm$     0.01&      29.03  $\pm$  0.05&      31.17  $\pm$     0.07 & 14.54 & 1474$\pm$46& dE5,N         & U7399A  & \\
 69 & 2019 & 1.15  $\pm$     0.01&      28.97 $\pm$   0.04&      31.16 $\pm$     0.07   & 14.55 & 1895$\pm$44& dE4,N         & I3735  & \\
 70 &   33 &  1.01  $\pm$     0.01&      28.57  $\pm$   0.05&      30.89 $\pm$     0.09  & 14.67 & 1093$\pm$52& d:E2,N:       & I3032  & \\
 71 &  200 &  1.16  $\pm$     0.01&      29.11  $\pm$  0.05&      31.30  $\pm$     0.07   & 14.69 &   16$\pm$20& dE2,N         &        &  \\
 72 &  571 &1.06 $\pm$  0.01&  29.61 $\pm$  0.08& 31.88 $\pm$  0.10  & 14.74 & 1047$\pm$37& SB0$_1$(6)    &        & 5 + dust \\
 73 &   21 &0.86 $\pm$ 0.01 &28.97 $\pm$ 0.04 &\nodata    & 14.75 &  486$\pm$25& dS0(4)        & I3025   &  6\\
 74 & 1488 &0.95 $\pm$ 0.01  & 28.57 $\pm$ 0.05 &\nodata   & 14.76 & 1079$\pm$25& E6:           & I3487  &  6\\
 75 & 1779 &0.90 $\pm$ 0.01 &28.42 $\pm$ 0.06  &\nodata   & 14.83 & 1313$\pm$45& dS0(6):       & I3612  &  6\\
 76 & 1895 &   1.12  $\pm$     0.01&      28.77  $\pm$  0.03&      31.00 $\pm$     0.06 & 14.91 & 1032$\pm$51& d:E6          & U7854   & \\
 77 & 1499 &0.67 $\pm$ 0.01  & 27.97 $\pm$ 0.07&\nodata        & 14.94 & -575$\pm$35& E3~pec~or~S0  & I3492   & 6 \\
 78 & 1545 & 1.25  $\pm$     0.01&      29.02  $\pm$  0.05&      31.13  $\pm$     0.07 & 14.96 & 2000$\pm$22& E4            & I3509 &  \\
 79 & 1192 &\nodata&\nodata &\nodata   & 15.04 & 1423$\pm$11& E3            & N4467  & 4  \\
 80 & 1857 &\nodata&\nodata &\nodata     & 15.07 &  634$\pm$69& dE4:,N?       & I3647   & 4 \\
 81 & 1075 &1.15 $\pm$     0.01&      28.85  $\pm$  0.08&      31.04  $\pm$     0.09  & 15.08 & 1844$\pm$40& dE4,N         & I3383  &  \\
 82 & 1948 &0.99 $\pm$     0.01 & 28.80 $\pm$     0.10&\nodata   & 15.10 & 1672$\pm$98& dE3           &        & 6 \\
 83 & 1627 &1.32 $\pm$     0.13& 28.96 $\pm$     0.05&  30.97  $\pm$ 0.27     & 15.16 &  236$\pm$41& E0            &        & 5 \\
 84 & 1440 & 1.19  $\pm$     0.01&      28.86  $\pm$   0.05&      31.02  $\pm$     0.07 & 15.20 &  382$\pm$21& E0            & I798   &  \\
 85 &  230 & 1.16  $\pm$     0.01&      29.06  $\pm$  0.09&      31.25  $\pm$      0.10  & 15.20 & 1429$\pm$20& dE4:,N:       & I3101  &  \\
 86 & 2050 & 1.11  $\pm$     0.01&      28.76  $\pm$  0.06&      30.99 $\pm$     0.08 & 15.20 & 1193$\pm$48& dE5:,N        & I3779  & \\
 87 & 1993 & 1.23  $\pm$     0.01&      28.97 $\pm$   0.03&      31.09  $\pm$     0.05 & 15.30 &  875$\pm$50& E0            &        &  \\
 88 &  751 & 1.25  $\pm$     0.01&      28.89  $\pm$     0.05&      30.99  $\pm$     0.06  & 15.30 &  697$\pm$36& dS0           & I3292  & \\
 89 & 1828 &1.18  $\pm$     0.01&      28.97  $\pm$   0.06&      31.13  $\pm$     0.08  & 15.33 & 1569$\pm$25& dE2,N         & I3635  & \\
 90 &  538 &1.11 $\pm$ 0.01  &29.57 $\pm$ 0.06 & 31.80 $\pm$ 0.09   & 15.40 &  620$\pm$37& E0            & N4309A  &\\
 91 & 1407 & 1.22  $\pm$     0.01&      28.99  $\pm$  0.03&      31.12  $\pm$     0.05 & 15.49 & 1001$\pm$11& dE2,N         & I3461&  \\
 92 & 1886 & 0.95 $\pm$     0.01 & 28.61 $\pm$ 0.14&\nodata    & 15.49 & 1159$\pm$65& dE5,N         &        & 6\\
 93 & 1199 &\nodata&\nodata&\nodata    & 15.50 &  1201$\pm$21& E2            &        & 4 \\
 94 & 1743 & 1.07  $\pm$     0.01&      28.96  $\pm$ 0.07&      31.23  $\pm$     0.09 & 15.50 & 1279$\pm$10& dE6           & I3602  & \\
 95 & 1539 & 1.15  $\pm$     0.01&      28.94  $\pm$   0.10&      31.14  $\pm$      0.11 & 15.68 & 1491$\pm$25& dE0,N         &        & \\
 96 & 1185 & 1.24  $\pm$     0.01&      29.03 $\pm$  0.09&      31.14  $\pm$     0.10  & 15.68 &  500$\pm$50& dE1,N         &        & \\
 97 & 1826 & 1.12  $\pm$     0.01&      28.83 $\pm$  0.06&      31.05 $\pm$     0.08  & 15.70 & 2033$\pm$38& dE2,N         & I3633  & \\
 98 & 1512 & 1.28  $\pm$  0.01 &   29.24 $\pm$   0.03   &  31.32 $\pm$   0.05      & 15.73 &  762$\pm$35& dS0~pec       &   &      \\
 99 & 1489 & 0.99 $\pm$     0.01&28.96 $\pm$ 0.09 &\nodata    & 15.89 &   80$\pm$50& dE5,N?        & I3490  & 6 \\
100 & 1661 & 1.26  $\pm$     0.01&      28.90 $\pm$   0.16&      31.00  $\pm$      0.16 & 15.97 & 1457$\pm$34& dE0,N         &        & \\
\enddata
\tablenotetext{a}{Notes:
(1) no SBF measurement possible (dust);
(2) no SBF measurement possible (edge-on disk or bar);
(3) very uncertain SBF measurement due to the precence of a bright star;
(4) no SBF measurement possible (sky subtraction)
(5) uncertain SBF measurement because of difficult sky or galaxy model subtraction;
(6) no SBF distance (galaxy too blue).}
\end{deluxetable}

\clearpage

\begin{deluxetable}{crcccccc}
\tabletypesize{\scriptsize}
\tablecaption{Comparison with the Previous SBF Distance Measurements\label{compare}}
\tablewidth{0pt}
\tablehead{
 &  &  & \multicolumn{4}{c}{$(m-M)$} \\
\cline{4-7} \\
\colhead{ID} &
\colhead{VCC} &
\colhead{Other Name} &
\colhead{ACSVCS} &
\colhead{NT00} &
\colhead{T01} &
\colhead{J04} \\ 
\colhead{} &
\colhead{} &
\colhead{} &
\colhead{(mag)} &
\colhead{(mag)} &
\colhead{(mag)} &
\colhead{(mag)}
}
\startdata
 1 & 1226&NGC4472&      31.17$\pm$0.07& 30.94$\pm$0.09  &   31.06$\pm$0.10 & \nodata \\
 2 & 1316&NGC4486&      31.18$\pm$0.07& 31.15$\pm$0.12  &   31.03$\pm$0.16 & \nodata \\
 3 & 1978&NGC4649&      31.19$\pm$0.07& 31.06$\pm$0.11  &   31.13$\pm$0.15 & \nodata \\
 4 &  881&NGC4406&      31.13$\pm$0.07& 31.45$\pm$0.12  &   31.17$\pm$0.14 & \nodata \\
 5 &  798&NGC4382&      31.26$\pm$0.05& \nodata         &   31.33$\pm$0.14 & \nodata \\
 6 &  763&NGC4374&      31.33$\pm$0.05& 31.17$\pm$0.10  &   31.32$\pm$0.11 & \nodata \\
 7 &  731&NGC4365&      31.84$\pm$0.06& 31.94$\pm$0.15  &   31.55$\pm$0.17 & \nodata \\
 9 & 1903&NGC4621&      30.87$\pm$0.06& 30.82$\pm$0.10  &   31.31$\pm$0.20 & \nodata \\
10 & 1632&NGC4552&      31.00$\pm$0.06& 31.00$\pm$0.10  &   30.93$\pm$0.14 & \nodata \\
11 & 1231&NGC4473&      30.92$\pm$0.06& 31.07$\pm$0.11  &   30.98$\pm$0.13 & \nodata \\
13 & 1154&NGC4459&      31.03$\pm$0.06& \nodata         &   31.04$\pm$0.22 & \nodata \\
15 & 2092&NGC4754&      31.04$\pm$0.06& \nodata         &   31.13$\pm$0.14 & \nodata \\
20 & 2000&NGC4660&      30.88$\pm$0.05& 31.30$\pm$0.16  &   30.54$\pm$0.19 & \nodata \\
22 & 1664&NGC4564&      31.00$\pm$0.06& \nodata         &   30.88$\pm$0.17 & \nodata \\
25 & 1938&NGC4638&      31.21$\pm$0.05& \nodata         &   31.68$\pm$0.26 & \nodata \\
26 & 1279&NGC4478&      31.15$\pm$0.06& 31.10$\pm$0.11  &   31.29$\pm$0.28 & \nodata \\
27 & 1720&NGC4578&      31.06$\pm$0.06& \nodata         &   31.34$\pm$0.13 & \nodata \\
29 & 1619&NGC4550&      30.95$\pm$0.05& 30.82$\pm$0.22  &   31.00$\pm$0.20 & \nodata \\
32 &  784&NGC4379&      31.00$\pm$0.06& \nodata         &   30.76$\pm$0.41 & \nodata \\
35 & 1321&NGC4489&      30.94$\pm$0.05& \nodata         &   31.26$\pm$0.15 & \nodata \\
36 &  828&NGC4387&      31.27$\pm$0.06& \nodata         &   31.65$\pm$0.73 & \nodata \\
37 & 1250&NGC4476&      31.23$\pm$0.07& 31.60$\pm$0.16  &   31.18$\pm$0.17 & \nodata \\
38 & 1630&NGC4551&      31.04$\pm$0.06& \nodata         &   31.19$\pm$0.17 & \nodata \\
39 & 1146&NGC4458&      31.07$\pm$0.05& 31.78$\pm$0.59  &   31.18$\pm$0.12 & \nodata \\
40 & 1025&NGC4434&      31.75$\pm$0.06& \nodata         &   32.14$\pm$0.17 & \nodata \\
48 & 1261&NGC4482&      31.29$\pm$0.06& \nodata         &   \nodata        &31.34$\pm$0.16 \\
   &     &       &                    & \nodata         &   \nodata        &31.74$\pm$0.21 \\
50 & 1422&IC3468 &      30.93$\pm$0.06& \nodata         &   \nodata        &31.64$\pm$0.18 \\
52 & 1871&IC3653 &      30.95$\pm$0.06& \nodata         &   30.84$\pm$0.45 & \nodata \\
53 &    9&IC3019 &      31.17$\pm$0.09& \nodata         &   \nodata        &31.00$\pm$0.10 \\
57 &  856&IC3328 &      31.13$\pm$0.07& \nodata         &   \nodata        &31.28$\pm$0.25 \\
59 & 1355&IC3442 &      31.14$\pm$0.08& \nodata         &   \nodata        &30.92$\pm$0.17 \\
60 & 1087&IC3381 &      31.11$\pm$0.06& \nodata         &   \nodata        &31.27$\pm$0.14 \\
   &     &       &                    & \nodata         &   \nodata        &31.39$\pm$0.18 \\
61 & 1297&NGC4486B&     31.06$\pm$0.07& 31.14$\pm$0.16  &   \nodata        & \nodata \\
\enddata
\end{deluxetable}

\clearpage
\begin{landscape}
\begin{deluxetable}{lrcccccccc}
\tabletypesize{\tiny}
\tablecaption{Mean Distances and Dispersions for Virgo Cluster Subsamples\label{mean}}
\tablewidth{0pt}
\tablehead{
\colhead{Sample} &
\colhead{$N$} &
\colhead{$\overline{{(m-M)}}$} &
\colhead{${\sigma(m-M)_{\rm o}}$} &
\colhead{$\overline{\sigma(m-M)}_{\rm m}$} &
\colhead{${\sigma(m-M)_{\rm i}}$} &
\colhead{$\overline{d}$} &
\colhead{${\sigma(d)}$} &
\colhead{$\overline{v}_{r}$} &
\colhead{${\sigma_{v_r}}$} \\
\colhead{} &
\colhead{} &
\colhead{(mag)} &
\colhead{(mag)} &
\colhead{(mag)} &
\colhead{(mag)} &
\colhead{(Mpc)} &
\colhead{(Mpc)} &
\colhead{(\kms)} &
\colhead{(\kms)}
}
\startdata
Giants                                                  & 54  & 31.134$\pm$0.031 & 0.227$\pm$0.040 & 0.065 & 0.212      & 16.9$\pm$0.2 & 1.7$\pm$0.3  & 1107$\pm$~72 & 528$\pm$53 \\
Dwarfs                                                  & 30  & 31.134$\pm$0.029 & 0.156$\pm$0.045 & 0.077 & 0.126      & 16.9$\pm$0.2 & 1.0$\pm$0.3  & 1180$\pm$~98 & 538$\pm$64 \\
$1.3 < (g_{475}-z_{850})_0 \le 1.6$                     & 39  & 31.124$\pm$0.032 & 0.201$\pm$0.045 & 0.066 & 0.183      & 16.8$\pm$0.2 & 1.4$\pm$0.3  & 1034$\pm$~87 & 541$\pm$69 \\
$1.0 \le (g_{475}-z_{850})_0 \le 1.3$                   & 45  & 31.143$\pm$0.031 & 0.208$\pm$0.044 & 0.073 & 0.188      & 16.9$\pm$0.2 & 1.5$\pm$0.3  & 1218$\pm$~76 & 512$\pm$45 \\
& & & & & & \\
Giants (no \wprime Cloud)                               & 50  & 31.079$\pm$0.017 & 0.120$\pm$0.011 & 0.064 & 0.088      & 16.4$\pm$0.1 & 0.7$\pm$0.1  & 1116$\pm$~77 & 546$\pm$55 \\
Dwarfs (no \wprime Cloud)                               & 29  & 31.114$\pm$0.021 & 0.112$\pm$0.013 & 0.076 & 0.065      & 16.7$\pm$0.2 & 0.5$\pm$0.1  & 1176$\pm$102 & 547$\pm$64 \\
$1.3 < (g$-$z)_0 \le 1.6$ (no \wprime Cloud)            & 37  & 31.088$\pm$0.021 & 0.127$\pm$0.011 & 0.064 & 0.098      & 16.5$\pm$0.2 & 0.7$\pm$0.1  & 1027$\pm$~91 & 555$\pm$70 \\
$1.0 \le (g$-$z)_0 \le 1.3$ (no \wprime Cloud)          & 42  & 31.096$\pm$0.017 & 0.110$\pm$0.012 & 0.066 & 0.072      & 16.6$\pm$0.1 & 0.6$\pm$0.1  & 1235$\pm$~80 & 521$\pm$47 \\
& & & & & & \\
Cluster A ($R \le 2^{\circ}$)                           & 32  & 31.119$\pm$0.020 & 0.113$\pm$0.012 & 0.071 & 0.072      & 16.7$\pm$0.2 & 0.6$\pm$0.1  & 1088$\pm$105 & 593$\pm$~68 \\
Cluster B ($R \le 1\fd5$)                               &  4  & 31.075$\pm$0.042 & 0.084$\pm$0.020 & 0.066 & 0.014      & 16.4$\pm$0.3 & 0.1$\pm$0.1  & ~958$\pm$111 & 222$\pm$~94 \\
\wprime Cloud                                           &  5  & 31.800$\pm$0.029 & 0.065$\pm$0.017 & 0.074 & $\le$0.065 & 22.9$\pm$0.3 & $\le$0.7     & 1042$\pm$113 & 253$\pm$141 \\
All Galaxies (no \wprime Cloud)                         & 79  & 31.092$\pm$0.013 & 0.118$\pm$0.008 & 0.069 & 0.082      & 16.5$\pm$0.1 & 0.6$\pm$0.1  & 1138$\pm$~61 & 544$\pm$~40 \\
\enddata
\end{deluxetable}
\clearpage
\end{landscape}

\clearpage

\begin{deluxetable}{llcccc}
\tabletypesize{\scriptsize}
\tablecaption{Axes of the Three-Dimensional Galaxy Distribution\label{shape}}
\tablewidth{0pt}
\tablehead{
\colhead{Sample} &
\colhead{$N$} &
\colhead{$a$} &
\colhead{$b$} &
\colhead{$c$} &
\colhead{$\Theta$} \\
\colhead{} &
\colhead{} &
\colhead{} &
\colhead{} &
\colhead{} &
\colhead{(deg)}  
}
\startdata
         {\it Unweighted} &&&&&\\
 & & & & & \\                                                  All galaxies &  84  &1&       0.43$\pm$       0.09 &       0.27$\pm$       0.06 &  12$\pm$   2  \\
                                             Gal. with $(m-M) < 31.5 $ &  79  &1&       0.74$\pm$       0.08 &       0.48$\pm$       0.05 &  29$\pm$  29  \\
                                           Giants with $(m-M) < 31.5 $ &  50  &1&       0.68$\pm$       0.09 &       0.41$\pm$       0.06 &  21$\pm$  21  \\
                                           Dwarfs with $(m-M) < 31.5 $ &  29  &1&       0.74$\pm$       0.12 &       0.42$\pm$       0.11 &  65$\pm$  30  \\
           Gal. with $(g_{475}-z_{850})_0 \le 1.3$ and $(m-M) < 31.5 $ &  42  &1&       0.79$\pm$       0.09 &       0.50$\pm$       0.09 &  54$\pm$  41  \\
             Gal. with $(g_{475}-z_{850})_0 > 1.3$ and $(m-M) < 31.5 $ &  37  &1&       0.62$\pm$       0.10 &       0.38$\pm$       0.06 &  24$\pm$  17  \\
                         Gal. with $M_B < 12$~mag  and $(m-M) < 31.5 $ &  17  &1&       0.55$\pm$       0.11 &       0.35$\pm$       0.09 &  30$\pm$  16  \\
Gal. with $M_B < 12$~mag, $(m-M) < 31.5 $  and  $ \delta (2000)> 10^o$ &  14  &1&       0.48$\pm$       0.12 &       0.25$\pm$       0.12 &  28$\pm$  15  \\

   Gal. WB &  15  &1&       0.28$\pm$       0.04 &       0.12$\pm$       0.05 &  12$\pm$   3  \\

                         Gal. with $M_B > 15$~mag  and $(m-M) < 31.5 $ &  15  &1&       0.73$\pm$       0.13 &       0.22$\pm$       0.11 &  69$\pm$  38  \\
                                      Gal. with $ \delta (2000)> 10^o$ &  67  &1&       0.69$\pm$       0.08 &       0.40$\pm$       0.06 &  26$\pm$  25  \\
      & & & & & \\                                                   
{\it Luminosity weighted} &&&&&\\
 & & & & & \\
                                                          All galaxies &  84  &1&       0.46$\pm$       0.16 &       0.26$\pm$       0.10 &  16$\pm$  15  \\
                                             Gal. with $(m-M) < 31.5 $ &  79  &1&       0.68$\pm$       0.12 &       0.41$\pm$       0.10 &  42$\pm$  33  \\
                                           Giants with $(m-M) < 31.5 $ &  50  &1&       0.65$\pm$       0.13 &       0.38$\pm$       0.09 &  33$\pm$  29  \\
                                           Dwarfs with $(m-M) < 31.5 $ &  29  &1&       0.70$\pm$       0.13 &       0.39$\pm$       0.11 &  63$\pm$  31  \\
           Gal. with $(g_{475}-z_{850})_0 \le 1.3$ and $(m-M) < 31.5 $ &  42  &1&       0.74$\pm$       0.11 &       0.46$\pm$       0.10 &  56$\pm$  36  \\
             Gal. with $(g_{475}-z_{850})_0 > 1.3$ and $(m-M) < 31.5 $ &  37  &1&       0.62$\pm$       0.13 &       0.35$\pm$       0.09 &  28$\pm$  23  \\
                         Gal. with $M_B < 12$~mag  and $(m-M) < 31.5 $ &  17  &1&       0.57$\pm$       0.14 &       0.33$\pm$       0.11 &  34$\pm$  23  \\
Gal. with $M_B < 12$~mag, $(m-M) < 31.5 $  and  $ \delta (2000)> 10^o$ &  14  &1&       0.50$\pm$       0.15 &       0.25$\pm$       0.12 &  31$\pm$  18  \\

 Gal. WB &  15  &1&       0.32$\pm$       0.08 &       0.10$\pm$       0.06 &  15$\pm$   5  \\

                         Gal. with $M_B > 15$~mag  and $(m-M) < 31.5 $ &  15  &1&       0.64$\pm$       0.15 &       0.25$\pm$       0.09 &  61$\pm$  31  \\
                                      Gal. with $ \delta (2000)> 10^o$ &  67  &1&       0.64$\pm$       0.13 &       0.36$\pm$       0.09 &  38$\pm$  30  \\
\enddata
\tablenotetext{~}{Note: The parameters {\it a, b,} and {\it c} are, respectively, the main, intermediate and minor axis of the distribution, normalized to the length of the main axis.  $N$ is the number of galaxies in each sample. The orientation angle, $\Theta$, is the angle between the line of sight and the main axis. The symbol $WB$ refers to a sample selected to be more similar to the West \& Blakeslee sample, with $B_T < 12.93$~mag, RA between 12.65 and 12.35 hrs and DEC between 10.5$^{\circ}$ and 14.5$^{\circ}$.}
\end{deluxetable}

\clearpage

\begin{deluxetable}{crccc}
\tabletypesize{\scriptsize}
\tablecaption{Distance Moduli for Different SBF Calibrations\label{tabledm}}
\tablewidth{0pt}
\tablehead{
\colhead{ID} &
\colhead{VCC} &
\colhead{$(m-M)_{bl}$} &
\colhead{$(m-M)_{l}$} &
\colhead{$(m-M)_{p}$} \\
\colhead{} &
\colhead{} &
\colhead{(mag)} &
\colhead{(mag)} &
\colhead{(mag)} 
}
\startdata
 1 & 1226 &     31.17 $\pm$      0.07 &     31.26 $\pm$      0.04 &     31.14 $\pm$      0.05  \\
 2 & 1316 &     31.17 $\pm$      0.07 &     31.27 $\pm$      0.04 &     31.12 $\pm$      0.06  \\
 3 & 1978 &     31.19 $\pm$      0.07 &     31.31 $\pm$      0.04 &     31.06 $\pm$      0.06  \\
 4 &  881 &     31.27 $\pm$      0.06 &     31.32 $\pm$      0.03 &     31.30 $\pm$      0.05  \\
 5 &  798 &     31.26 $\pm$      0.05 &     31.23 $\pm$      0.03 &     31.27 $\pm$      0.05  \\
 6 &  763 &     31.33 $\pm$      0.06 &     31.36 $\pm$      0.03 &     31.37 $\pm$      0.05  \\
 7 &  731 &     31.84 $\pm$      0.06 &     31.91 $\pm$      0.04 &     31.84 $\pm$      0.05  \\
 9 & 1903 &     30.86 $\pm$      0.06 &     30.92 $\pm$      0.03 &     30.89 $\pm$      0.05  \\
10 & 1632 &     31.05 $\pm$      0.06 &     31.12 $\pm$      0.04 &     31.05 $\pm$      0.05  \\
11 & 1231 &     30.92 $\pm$      0.06 &     30.96 $\pm$      0.04 &     30.95 $\pm$      0.05  \\
13 & 1154 &     31.03 $\pm$      0.06 &     31.08 $\pm$      0.04 &     31.06 $\pm$      0.05  \\
14 & 1062 &     30.92 $\pm$      0.06 &     30.96 $\pm$      0.04 &     30.96 $\pm$      0.05  \\
15 & 2092 &     31.04 $\pm$      0.06 &     31.09 $\pm$      0.04 &     31.06 $\pm$      0.05  \\
16 &  369 &     31.00 $\pm$      0.06 &     31.03 $\pm$      0.04 &     31.03 $\pm$      0.05  \\
17 &  759 &     31.15 $\pm$      0.06 &     31.20 $\pm$      0.04 &     31.17 $\pm$      0.05  \\
18 & 1692 &     31.16 $\pm$      0.06 &     31.19 $\pm$      0.04 &     31.19 $\pm$      0.05  \\
19 & 1030 &     31.12 $\pm$      0.05 &     31.07 $\pm$      0.03 &     31.10 $\pm$      0.05  \\
20 & 2000 &     30.88 $\pm$      0.05 &     30.84 $\pm$      0.04 &     30.88 $\pm$      0.05  \\
22 & 1664 &     31.00 $\pm$      0.06 &     31.03 $\pm$      0.04 &     31.04 $\pm$      0.05  \\
24 &  944 &     31.02 $\pm$      0.05 &     31.01 $\pm$      0.03 &     31.04 $\pm$      0.05  \\
25 & 1938 &     31.21 $\pm$      0.05 &     31.15 $\pm$      0.04 &     31.18 $\pm$      0.05  \\
26 & 1279 &     31.15 $\pm$      0.06 &     31.16 $\pm$      0.04 &     31.18 $\pm$      0.05  \\
27 & 1720 &     31.06 $\pm$      0.06 &     31.08 $\pm$      0.04 &     31.09 $\pm$      0.05  \\
28 &  355 &     30.94 $\pm$      0.06 &     30.95 $\pm$      0.04 &     30.97 $\pm$      0.05  \\
29 & 1619 &     30.95 $\pm$      0.05 &     30.89 $\pm$      0.03 &     30.92 $\pm$      0.05  \\
30 & 1883 &     31.10 $\pm$      0.05 &     31.05 $\pm$      0.04 &     31.07 $\pm$      0.05  \\
31 & 1242 &     30.96 $\pm$      0.07 &     30.91 $\pm$      0.05 &     30.94 $\pm$      0.06  \\
32 &  784 &     31.00 $\pm$      0.06 &     30.98 $\pm$      0.04 &     31.01 $\pm$      0.05  \\
33 & 1537 &     31.00 $\pm$      0.05 &     30.94 $\pm$      0.04 &     30.97 $\pm$      0.05  \\
34 &  778 &     31.25 $\pm$      0.06 &     31.20 $\pm$      0.04 &     31.23 $\pm$      0.05  \\
35 & 1321 &     30.94 $\pm$      0.05 &     30.89 $\pm$      0.04 &     30.91 $\pm$      0.05  \\
36 &  828 &     31.27 $\pm$      0.06 &     31.26 $\pm$      0.04 &     31.29 $\pm$      0.05  \\
37 & 1250 &     31.23 $\pm$      0.07 &     31.23 $\pm$      0.06 &     31.22 $\pm$      0.07  \\
38 & 1630 &     31.04 $\pm$      0.06 &     31.06 $\pm$      0.04 &     31.07 $\pm$      0.05  \\
39 & 1146 &     31.07 $\pm$      0.05 &     31.02 $\pm$      0.04 &     31.05 $\pm$      0.05  \\
40 & 1025 &     31.75 $\pm$      0.06 &     31.75 $\pm$      0.04 &     31.77 $\pm$      0.05  \\
41 & 1303 &     31.12 $\pm$      0.06 &     31.09 $\pm$      0.04 &     31.13 $\pm$      0.05  \\
42 & 1913 &     31.20 $\pm$      0.06 &     31.16 $\pm$      0.04 &     31.19 $\pm$      0.05  \\
43 & 1327 &     31.27 $\pm$      0.07 &     31.28 $\pm$      0.05 &     31.30 $\pm$      0.06  \\
45 & 1475 &     31.10 $\pm$      0.06 &     31.08 $\pm$      0.04 &     31.09 $\pm$      0.05  \\
46 & 1178 &     30.99 $\pm$      0.06 &     30.98 $\pm$      0.04 &     31.01 $\pm$      0.05  \\
47 & 1283 &     31.19 $\pm$      0.06 &     31.19 $\pm$      0.04 &     31.22 $\pm$      0.05  \\
48 & 1261 &     31.29 $\pm$      0.06 &     31.30 $\pm$      0.05 &     31.29 $\pm$      0.06  \\
49 &  698 &     31.36 $\pm$      0.05 &     31.30 $\pm$      0.04 &     31.33 $\pm$      0.05  \\
50 & 1422 &     30.93 $\pm$      0.06 &     30.91 $\pm$      0.05 &     30.92 $\pm$      0.06  \\
52 & 1871 &     30.95 $\pm$      0.06 &     30.93 $\pm$      0.05 &     30.96 $\pm$      0.06  \\
53 &    9 &     31.17 $\pm$      0.09 &     31.20 $\pm$      0.08 &     31.18 $\pm$      0.08  \\
54 &  575 &     31.72 $\pm$      0.06 &     31.68 $\pm$      0.05 &     31.70 $\pm$      0.06  \\
55 & 1910 &     31.03 $\pm$      0.06 &     30.99 $\pm$      0.04 &     31.02 $\pm$      0.05  \\
56 & 1049 &     31.02 $\pm$      0.08 &     31.06 $\pm$      0.06 &     31.03 $\pm$      0.07  \\
57 &  856 &     31.13 $\pm$      0.07 &     31.12 $\pm$      0.05 &     31.12 $\pm$      0.06  \\
58 &  140 &     31.07 $\pm$      0.07 &     31.08 $\pm$      0.05 &     31.07 $\pm$      0.06  \\
59 & 1355 &     31.14 $\pm$      0.08 &     31.15 $\pm$      0.07 &     31.14 $\pm$      0.08  \\
60 & 1087 &     31.11 $\pm$      0.06 &     31.07 $\pm$      0.05 &     31.09 $\pm$      0.06  \\
61 & 1297 &     31.06 $\pm$      0.07 &     31.11 $\pm$      0.05 &     31.09 $\pm$      0.06  \\
62 & 1861 &     31.04 $\pm$      0.06 &     30.99 $\pm$      0.05 &     31.01 $\pm$      0.06  \\
63 &  543 &     30.98 $\pm$      0.07 &     30.97 $\pm$      0.05 &     30.97 $\pm$      0.06  \\
64 & 1431 &     31.04 $\pm$      0.06 &     30.98 $\pm$      0.05 &     31.01 $\pm$      0.06  \\
65 & 1528 &     31.06 $\pm$      0.06 &     31.03 $\pm$      0.05 &     31.04 $\pm$      0.06  \\
66 & 1695 &     31.09 $\pm$      0.08 &     31.13 $\pm$      0.06 &     31.10 $\pm$      0.07  \\
67 & 1833 &     31.05 $\pm$      0.07 &     31.05 $\pm$      0.05 &     31.05 $\pm$      0.06  \\
68 &  437 &     31.17 $\pm$      0.07 &     31.15 $\pm$      0.06 &     31.16 $\pm$      0.06  \\
69 & 2019 &     31.16 $\pm$      0.07 &     31.16 $\pm$      0.05 &     31.16 $\pm$      0.06  \\
70 &   33 &     30.89 $\pm$      0.09 &     30.94 $\pm$      0.06 &     30.90 $\pm$      0.08  \\
71 &  200 &     31.30 $\pm$      0.07 &     31.29 $\pm$      0.06 &     31.29 $\pm$      0.07  \\
72 &  571 &     31.88 $\pm$      0.10 &     31.92 $\pm$      0.09 &     31.89 $\pm$      0.09  \\
76 & 1895 &     31.00 $\pm$      0.06 &     31.01 $\pm$      0.04 &     31.00 $\pm$      0.05  \\
78 & 1545 &     31.13 $\pm$      0.07 &     31.08 $\pm$      0.06 &     31.10 $\pm$      0.07  \\
81 & 1075 &     31.04 $\pm$      0.09 &     31.04 $\pm$      0.08 &     31.04 $\pm$      0.09  \\
83 & 1627 &     30.78 $\pm$      0.06 &     30.81 $\pm$      0.04 &     30.82 $\pm$      0.05  \\
84 & 1440 &     31.02 $\pm$      0.07 &     31.01 $\pm$      0.06 &     31.01 $\pm$      0.06  \\
85 &  230 &     31.25 $\pm$      0.10 &     31.25 $\pm$      0.10 &     31.25 $\pm$      0.10  \\
86 & 2050 &     30.99 $\pm$      0.08 &     31.01 $\pm$      0.07 &     31.00 $\pm$      0.07  \\
88 &  751 &     30.99 $\pm$      0.06 &     30.95 $\pm$      0.05 &     30.97 $\pm$      0.06  \\
87 & 1993 &     31.09 $\pm$      0.05 &     31.06 $\pm$      0.04 &     31.07 $\pm$      0.05  \\
89 & 1828 &     31.14 $\pm$      0.08 &     31.12 $\pm$      0.07 &     31.12 $\pm$      0.07  \\
90 &  538 &     31.80 $\pm$      0.09 &     31.82 $\pm$      0.07 &     31.81 $\pm$      0.08  \\
91 & 1407 &     31.12 $\pm$      0.05 &     31.09 $\pm$      0.04 &     31.10 $\pm$      0.05  \\
94 & 1743 &     31.23 $\pm$      0.09 &     31.26 $\pm$      0.08 &     31.24 $\pm$      0.09  \\
95 & 1539 &     31.14 $\pm$      0.12 &     31.14 $\pm$      0.11 &     31.13 $\pm$      0.11  \\
96 & 1185 &     31.14 $\pm$      0.10 &     31.11 $\pm$      0.09 &     31.12 $\pm$      0.10  \\
97 & 1826 &     31.05 $\pm$      0.08 &     31.06 $\pm$      0.07 &     31.05 $\pm$      0.08  \\
98 & 1512 &     31.32 $\pm$      0.05 &     31.27 $\pm$      0.04 &     31.30 $\pm$      0.05  \\
100& 1661 &     31.00 $\pm$      0.16 &     30.96 $\pm$      0.16 &     30.98 $\pm$      0.16  \\
\enddata
\tablecomments{Distance moduli: $(m-M)_{bl}$ = broken linear relation; 
$(m-M)_{l}$ = single linear relation;
$(m-M)_{p}$ = polynomial relation.}
\end{deluxetable}

\begin{deluxetable}{lccccc}
\tablecaption{Comparison of SBF Calibrations\label{cali}}
\tablewidth{0pt}
\tablehead{
\colhead{Relation} &
\colhead{$N$} &
\colhead{$\chi^{2(a)}_r$} &
\colhead{$\chi^{2(a)}_{d}$} &
\colhead{$\sigma^{(b)}$} &
\colhead{$|(m-M)_{bl}-(m-M)|^{(c)}$} \\
\colhead{} &
\colhead{} &
\colhead{} &
\colhead{} &
\colhead{(mag)} &
\colhead{(mag)}
}
\startdata
Broken linear&79&28.5&0.91&0.12& \nodata\\
Linear&79&36.7&0.99&0.12&$0.03\pm 0.02$\\
Polynomial&79&28.1&0.91&0.12&$0.02\pm 0.02$\\
Broken linear $(g_{475}-z_{850})_0 \le 1.5$&76&27.3&0.93&0.12& \nodata\\
Linear $(g_{475}-z_{850})_0 \le 1.5$ &76&27.1&0.93&0.12&$0.03\pm 0.02$ \\
Polynomial  $(g_{475}-z_{850})_0 \le 1.5$ &76&28.9&0.95&0.12&$0.02\pm 0.01$\\
\enddata

\tablenotetext{a}{$\chi^2_r$ and $\chi^2_d$ are the reduced $\chi^2$ of the fits and the reduced $\chi^2$ when taking into account the uncertainties due to depth effects
 in the Virgo Cluster (0.11~mag) and the ``cosmic" dispersion in $\overline M_{850}$ (0.05~mag).}
\tablenotetext{b}{$rms$ scatter around the fitted relation.}
\tablenotetext{c}{Average absolute difference relative to the broken linear ($bl$) relation.}
\end{deluxetable}

\end{document}